\documentclass[structabstract
]{aa}  

\usepackage{graphicx,natbib}
\usepackage{txfonts}
\usepackage{dcolumn}
\usepackage{lscape}

\newcommand{\um}{\,\mu\rm{m}}

\newcommand{\degree}{\symbol{23}}

\newcommand{\lastfootnote}{\footnotemark[\value{footnote}]}

\begin{document}
   \title{Observing planet-disk interaction in debris disks}

   \author{Steve Ertel\inst{1,2}
        \and
          Sebastian Wolf\inst{2}
        \and
          Jens Rodmann\inst{3,4}
   }
   \institute{
          Universit\'e Joseph Fourier/CNRS, Laboratoire d'Astrophysique de Grenoble, UMR 5571, Grenoble, France\\
          \email{steve.ertel@obf.ujf-grenoble.fr} 
        \and
          Christian-Albrechts-Universit\"at zu Kiel, Institut f\"ur Theoretische Physik und Astrophysik, Leibnizstra{\ss}e 15, 24098 Kiel, Germany
        \and
          European Space Agency, Space Environment and Effects Section, Keplerlaan 1, PO Box 299, NL-2200 AG Noordwijk, The Netherlands
        \and
          Georg-August-Universit\"at G\"ottingen, Institut f\"ur Astrophysik, Friedrich-Hund-Platz 1, 37077 G\"ottingen, Germany
   }
   \date{}

  \abstract
   {Debris disks are commonly considered to be a by-product of planet formation. Structures in debris disks induced by planet-disk interaction are promising to provide valuable constraints on the existence and properties of embedded planets.}
   {We investigate the observability of structures in debris disks induced by planet-disk interaction with future facilities in a systematic way. High-sensitivity, high angular resolution observations with large \mbox{(sub-)mm} interferometers and large space-based telescopes operating in the near- to mid-infrared wavelength range are considered.}
   {The observability of debris disks with the Atacama Large Millimeter/submillimeter Array (ALMA) is studied on the basis of a simple analytical disk model. Furthermore, $N$-body simulations are used to model the spatial dust distribution in debris disks under the influence of planet-disk interaction. From these simulations, images at optical scattered light to millimeter thermal re-emission are computed. Available information about the expected capabilities of ALMA and the {\it James Webb} \normalfont Space Telescope (JWST) are used to investigate the observability of characteristic disk structures with these facilities through spatially resolved imaging.}
   {Our simulations show that planet-disk interaction can result in prominent structures in the whole considered wavelength range. The exact result depends on the configuration of the planet-disk system and on the observing wavelength which provides the opportunity of detecting and characterizing extrasolar planets in a range of masses and radial distances from the star that is not accessible to other techniques. Facilities that will be available in the near future at both considered wavelength ranges are shown to provide the capabilities to spatially resolve and characterize structures in debris disks that arise because of planet-disk interaction. Limitations are revealed and suggestions for possible instrument setups and observing strategies are given. In particular, ALMA is limited by its sensitivity to surface brightness, which requires a trade-off between sensitivity and spatial resolution. Space-based mid-infrared observations will be able to detect and spatially resolve regions in debris disks even at a distance of several tens of AU from the star, where the emission from debris disks in this wavelength range is expected to be low.}
   {Both ALMA and the planned space-based near- to mid-infrared telescopes will provide unprecedented capabilities to study planet-disk interaction in debris disks. In particular, a combination of observations at both wavelengths will provide very strong constraints on the planetary/planetesimal systems.}


   \maketitle

\section{Introduction}
\label{intro}

The dust detected in debris disks is thought to be removed from those systems by the stellar radiation on time scales that are short compared to their ages. This means that the dust must be transient, or more likely continuously replenished by ongoing collisions of bigger objects such as planetesimals left over from the planet formation process \citep[for a recent review see, e.g.,][]{kri10}. The presence of planets and debris disks is hence thought to be correlated. In a system with a debris disk and one or more planets, one would expect gravitational interaction between the dust grains and the planet, trapping them into resonance \citep{wya06, wol07, sta08, sta09b}. This results in structures in the disk that may be observable, which in turn provides a method to infer and characterize planets that are in a regime of masses, brightnesses, and radial distances from the star that is not accessible through other techniques such as radial velocity measurements or direct imaging \citep[e.g.,][]{udr08,mar08,kal08}.

Clumpy structures in debris disks have been observed in several cases (e.g., $\epsilon$\,Eri, \citealt{gre98}; AU\,Mic, \citealt{liu04}; HD\,107146, \citealt{cor09, hug11}). In this work, we investigate the observability of structures in debris disks in a systematic way. We set up several initial conditions for the planetary mass and orbit as well as for the dust distribution in the disk from which we simulate the spatial dust distribution using $N$-body simulations. Our results from the dynamical simulations are consistent with earlier works \citep[e.g.,][]{hol03,wya06,sta08}. In contrast to these works which focused mostly on the phenomenological and theoretical description of the structures as well as on the processes of planet-disk interaction, the goals of the present work are the evaluation of the observability and the development of strategies for observations of planet disk interaction. Thus, we use the $N$-body simulations as a tool to produce realistic structures in debris disks, but discuss these structures only briefly emphasizing on the consequences of different planet-disk configurations on the observability. To make predictions on the feasibility to detect and spatially resolve characteristic structures, available information on the capabilities of the Atacama Large Millimeter/submillimeter Array (ALMA) and {\it James Webb} Space Telescope (JWST) are used representative for near-future facilities.

We describe the approach used in our $N$-body simulations and for the image generation in Sect.~\ref{modust}. The results from the dynamical simulations are presented and briefly discussed in Sect.~\ref{dyn_res}. In Sects.~\ref{obs_ALMA} to~\ref{obs_JWST}, we evaluate how large-scale disk structures may be observed with different facilities. Conclusions are drawn in Sect~\ref{conc}.

\section{Modeling planet-disk interaction in debris disks}
\label{modust}

In this section, we describe the initial conditions employed for modeling planet-disk interaction. We simulate the dynamical evolution of a large ensemble of dust particles in the gravitational potential of a central star under the influences of planetary perturbations, radiation pressure, Poynting-Robertson effect, and stellar-wind drag. For these simulations, the code {\ttfamily MODUST} \citep{rod06} is employed. Electromagnetic (Lorentz) forces on charged dust particles are neglected \citep{gus94,hol03}. Mutual collisions of dust particles are not considered in the dynamical modeling. A brief description of the methods and numerical approach to simulate the dynamical evolution of the dust grains as well as of the approach to simulate images from the resulting density distributions can be found in the appendix.

\subsection{Initial conditions}
\label{sect_initial}

We use several initial conditions (planetary orbit and mass as well as initial dust distribution) selected by analogy to our solar system and to other known debris disks in order to study the structures that can be expected depending on the configuration of the respective system. Since the simulation of the resulting dust distribution is very time-consuming, we are not able to perform a dense sampling of a large, high dimensional parameter space. Therefore, we focus on the following parameter regime:

\begin{itemize}
 \item One central star (no multiple systems),
 \item a solar-type star as central star, photospheric emission realized by a black body with $T_{\rm eff} = 5778\,\rm K$, $L_\star = 1.0\,{\rm L_\odot}$, $M_\star = 1.0\,{\rm M_\odot}$, $\xi = 0.35$ \citep{gus94},
 \item one planet (considered to dominate the dynamics of the system),
 \item one initial dust disk (no initial multi-ring systems),
 \item planetary orbit and initial dust disk are coplanar.
\end{itemize}

Astronomical silicate with a bulk density of $2.7\,{\rm g/cm^3}$ \citep{dra84, wei01} is employed for the chemical composition of the dust. In the following, the initial conditions for the dust distribution (spatial distribution and grain size distribution) and for the planetary orbit are described and motivated.

\subsubsection{Treatment of dust creation}

We assume the dust to be produced through collisions in a disk of planetesimals. It is then redistributed because of the effects of stellar radiation and wind, and the gravitational interaction with the planet (Eq.~\ref{eq_forces}). This is realized by placing the initial dust distribution at the same position as the planetesimal disk postulated to produce the dust. Particles that are lost because they are sublimated or ejected from the system are replaced by new ones from this reservoir of initial dust grains. Furthermore, the grain size distribution of {\it all grains in the system} applied follows a power-law with exponent $-3.5$ as expected from an equilibrium collisional cascade \citep{doh69}. This approach is used to mimic dust production through collisions of the parent bodies. Note that the local grain size distribution may be significantly different due to the redistribution of the grains through dynamical interaction with the planet as well as through the effect of radiation pressure and Poynting-Robertson drag \citep{mor02,mor03}. 

\subsubsection{Grain size distribution}

The grain size can be expressed by the corresponding value of $\beta$. The correlation between these two quantities is given by
\begin{equation}
\label{beta}
  \beta = \frac{3 L_\star}{16 \pi c G M_\star} \frac{Q_{\rm pr}}{\rho a} \approx 575 \left(\frac{\rho}{\rm kg\,m^{-3}}\right)^{-1} \left(\frac{a}{\um}\right)^{-1} \left(\frac{L_\star}{\rm L_\odot}\right) \left(\frac{M_\star}{\rm M_\odot}\right)^{-1},
\end{equation}
where $a$ is the grain size, $Q_{\rm pr}$ is the efficiency of radiation pressure on a grain, $L_\star$ and $M_\star$ are the luminosity and mass of the star, $c$ is the speed of light, $G$ is the gravitational constant, and $\rho$ is the bulk density of the dust grains which are assumed to be spherical and compact. The approximation was given by \citet{wya99} and is valid in the geometrical optics approximation. Since the critical parameter for the computations is $\beta$, it is used in {\tt MODUST} directly to describe the grain properties (instead of grain size and optical properties). Around a solar-type star, the approximation gives good results for $\beta \leq 0.5$ \citep{rod06}. The lower boundary of the applied grain size distribution is defined by $\beta = 0.5$ ($a \approx 0.43\um$), because smaller grains (higher values of $\beta$) are ejected from the system by radiation pressure. The upper boundary of the distribution is set to a grain size of 2\,mm ($\beta \approx 1.06\times10^{-4}$ for the applied chemical composition). The contribution of larger grains to the simulated observations is neglected.

The grain size distribution is sampled with 50 grain sizes between the lower and the upper grain size, distributed logarithmically (to properly sample the thermal re-emission of different grain sizes for the final images). For each size bin, the dust spatial distribution is simulated. A total number of 1000 test particles is used in each size bin. The system evolves over five times the corresponding Poynting-Robertson time scale of grains of the corresponding $\beta$ at the outer edge of the disk, but not longer than the assumed age of the system (see motivation of the single runs, Sect.~\ref{sect_paraspace}). At the end of each run, 100 snapshots of the particle distribution \citep{rod06} are taken in the reference frame corotating with the planet to virtually increase the number of particles in each size bin. The snapshots are equally distributed over the last 10\% of each run time (see description of the different runs, Sect.~\ref{sect_paraspace}).

\subsubsection{Spatial dust distribution}

Gaussian distributions are applied to set the initial eccentricities and inclinations of the test particles. At the start of the integration, the majority of dust grains have orbital eccentricities between 0~and~$\sim 0.4$ (rms of the Gaussian distribution = 0.15). The full width at half maximum of the inclination distribution is set to $10\degree$. These values are consistent with measurements on dust parent bodies in the Kuiper Belt of the solar system \citep{jew96, vit10}. The semi-major axes of the particles are distributed following a power-law \citep{wol03} from a lowest to a highest value considered as the inner and outer radius of the disk.

\subsubsection{Explored parameter space}
\label{sect_paraspace}

The parameter space explored is motivated by analogy to our solar system and to other known debris disks. We use the position and shape of the dust disk as well as the position, mass, and eccentricity of the planet as parameters for our initial conditions. The different combinations of values explored for these parameters are listed in Table~\ref{tab_initial} and described and motivated below.

\smallskip
\noindent\emph{A first sequence of initial conditions (Ia to Id)} places the planet in 1:1 or in 2:1 resonance with the initial planetesimal belt. The initial dust disk is a narrow ring ($R_{\rm out} = 1.1\,R_{\rm in}$) with constant surface density and two realizations of $R_{\rm in}$ (5\,AU for runs Ia and Ib and 50\,AU for runs Ic and Id). The planet has a mass of $1\,\rm M_J$ and is on a circular orbit. The two assumed positions of the dust ring are motivated by our solar system (Asteroid Belt and Kuiper Belt) and typical positions of the dust in other known debris disks around Sun-like stars (e.g., $\epsilon$\,Eri, \citealt{bac09}; HD\,105). An age (run time of the simulation) of 50\,Myr is assumed for these runs \citep[e.g., the approximate age of HD\,105;][]{apa08}.

\smallskip
\noindent\emph{The second sequence of initial conditions (IIa to IId)} is motivated by the HD\,107146 debris disk \citep{ard04,ert11}. This disk is particularly well-suited for studying structures in debris disks because of its nearly face-on orientation and the large amount of complementary data available. Furthermore, \mbox{(sub-)mm} observations of the disk have been published by \citet{cor09} and \citet{hug11}, showing clumpy structures in the surface brightness that could be the signposts of planet-disk interaction\footnote{However, it is doubtful whether these structures are real, since they are only visible at a signal-to-noise ratio (S/N) of $\approx 5$ to 6 (2 to 3 above the average surface brightness of the disk) and their position in the two observations by \citet{cor09} and \citet{hug11} seems to be inconsistent.}. Here, we employ a set of initial conditions where the planet is orbiting at the inner edge of an extended disk with a large inner hole. The inner and outer radius of the disk are set to $R_{\rm in} = 70\,{\rm AU}$ and $R_{\rm out} = 250\,{\rm AU}$. The radial surface density distribution is described by a power-law with an index of $\gamma = -0.5$ \citep[e.g.,][]{wol03}. This model is similar to the results of detailed multi-wavelength modeling of the HD\,107146 debris disk \citep{ert11}, although a more complex radial density distribution has been found there. However, such a complex distribution might be the result of the interaction of the disk with a possible planet and the temporal evolution of the disk. We explore different parameters for the planet. It is on a circular orbit and has a mass of $0.5\,{\rm M_J}$, $1.0\,{\rm M_J}$, and $5.0\,{\rm M_J}$ in the runs IIa, IIb, and IIc. It is on an eccentric orbit ($e = 0.1$) and has a mass of $1.0\,{\rm M_J}$ in run IId. The age of the system is increased compared to that employed in our sequence I to account for the slightly higher age of HD\,107146 \citep[e.g.,][]{roc09}. An age of 100\,Myr is employed.

\smallskip
\noindent\emph{A third sequence of models (IIIa to IIId)} places the planet within a broad disk. Therefore we modify the initial conditions of our sequence II, so that the disk is now closer to the star, ranging from $R_{\rm in} = 35\,{\rm AU}$ to $R_{\rm out} = 210\,{\rm AU}$. The inner radius is comparable to that of the Kuiper Belt. The power-law index of the radial dust distribution and the parameters of the planet remain the same.

\begin{table}
\caption{Initial conditions for the different runs with {\ttfamily MODUST}}
\label{tab_initial}
\begin{center}
\begin{tabular}{ccccccc}
\hline\hline                 
 Run & $R_{\rm in}$ [AU] & $R_{\rm out}$ [AU] & $\alpha$ & $M_{\rm pl}$ [$\rm M_J$] & $a_{\rm pl}$ [AU] & $e_{\rm pl}$   \\
\hline
 Ia   & 5.0              & 5.5                & 0.0      & 1.0             & 5.0      & 0.0 \\
 Ib   & 5.0              & 5.5                & 0.0      & 1.0             & 3.15     & 0.0 \\
 Ic   & 50.0             & 55.0               & 0.0      & 1.0             & 50.0     & 0.0 \\
 Id   & 50.0             & 55.0               & 0.0      & 1.0             & 31.5     & 0.0 \\
\hline
 IIa  & 70.0             & 250.0              & $-0.5$   & 0.5             & 70.0     & 0.0 \\
 IIb  & 70.0             & 250.0              & $-0.5$   & 1.0             & 70.0     & 0.0 \\
 IIc  & 70.0             & 250.0              & $-0.5$   & 5.0             & 70.0     & 0.0 \\
 IId  & 70.0             & 250.0              & $-0.5$   & 1.0             & 70.0     & 0.1 \\
\hline
 IIIa & 35.0             & 210.0              & $-0.5$   & 0.5             & 70.0     & 0.0 \\
 IIIb & 35.0             & 210.0              & $-0.5$   & 1.0             & 70.0     & 0.0 \\
 IIIc & 35.0             & 210.0              & $-0.5$   & 5.0             & 70.0     & 0.0 \\
 IIId & 35.0             & 210.0              & $-0.5$   & 1.0             & 70.0     & 0.1 \\
\hline
\end{tabular}
\end{center}
\end{table}

\subsection{Image creation}

We employ Mie theory to compute the optical properties of the dust using the tool {\ttfamily miex} \citep{wol04}. To create images from the derived dust distributions, we compute the thermal re-emission and the scattered stellar light for each dust particle at a set of observing wavelengths. The fluxes are then projected onto the plane of the sky considering the scattering phases. This can be performed for arbitrary orientations of the disk (assuming an optically thin debris disk). The image extension in one dimension used is 301 pixels for sequence~I (resulting in a pixel resolution of 0.1\,AU for runs Ia and Ib and 1.0\,AU for runs Ic and Id) and 251 pixels for sequence~II and sequence~III (resulting in a pixel resolution of 2.0\,AU). The images for different grain sizes are then weighted following the grain size distribution and added to compose a final image for each model at each wavelength.

\subsection{Comparison to previous simulations}

\begin{figure*}
\centering
\includegraphics[width=1\linewidth]{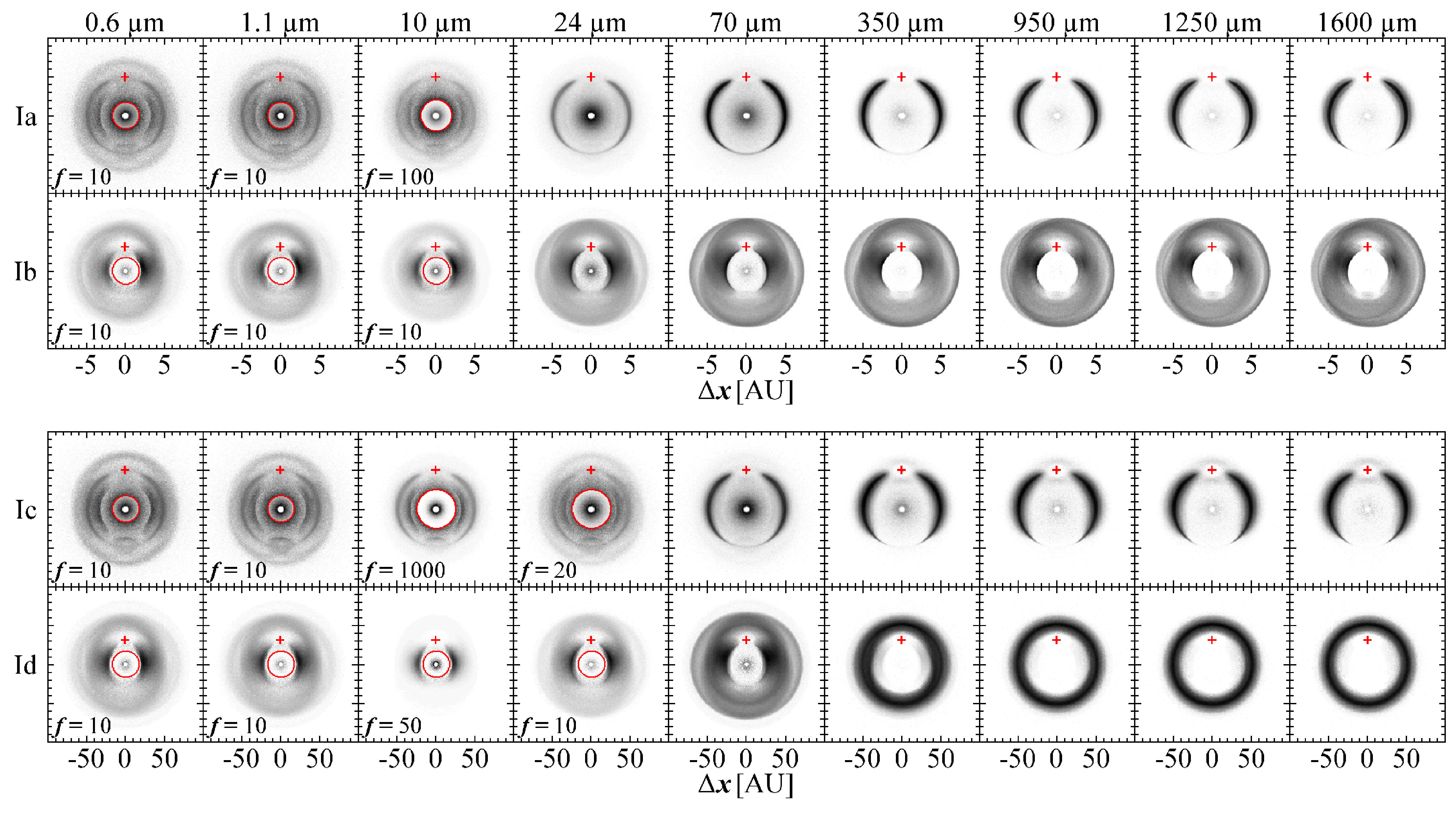}
\caption{Simulated images from model sequence I for a face-on orientation of the disk at different wavelengths. The wavelength is indicated at the top of each column. The star is at the center of each image. The position of the planet is indicated by the red cross in each image. The flux is given in arbitrary units and displayed in a logarithmic stretch from zero to peak value. The region marked by a red circle in some images has been attenuated by a factor $f$ to properly display the dynamical range of the images. The value of $f$ is then indicated in the lower-left corner of each image. If no value of $f$ is given, no attenuation has been applied. Note that a disk without a planet would show no azimuthal structures, but a smooth, featureless radial surface brightness distribution (besides inner and outer disk radius).}
\label{resonances_fo}%
\end{figure*}

In the images produced from our simulations (Fig.~\ref{resonances_fo} and~\ref{107146_fo}), it is possible to identify the structures described by \citet{kuc03} using simple geometrical arguments. The general goal and approach of our simulations is similar to those presented by \citet{sta08}. However, our models cover a different region of the parameter space (more massive planets, farther away from the star, larger dust grains with smaller $\beta$). While \citet{sta08} concentrated on the effects of terrestrial planets on exozodiacal dust clouds, the scope of the present work is to search for structures induced by giant planets in debris disks. Furthermore, we attempt to simulate observations of the structures found by our modeling. This requires a good sampling of the grain size dependent emissivity of the dust and, consequently, of the grain size distribution. The grains in \citet{sta08} are assumed to be no larger than $\sim 120\um$ and to be produced by an external source (i.e., a planetesimal belt at considerable distance from the modeled region), which is a reasonable assumption for exozodiacal dust. In the present case, the dust-producing planetesimal belt has to be taken into account. Accordingly, larger grains have to be considered. Furthermore, these grains cannot be neglected, since we attempt to simulate observations at \mbox{(sub-)mm} wavelengths, where millimeter-sized dust grains significantly contribute to the dust emission. Thus, a total of 50 grain size bins is used in the present work in contrast to the five bins in \citet{sta08}, and we consider grains up to a size of 2\,mm.

Nonetheless, the results from both studies can be compared qualitatively. In particular, for models that reproduce similar parameter regimes the results are very similar and differences can be attributed to the above described differences in the approaches and to the differences in the nature of the images themselves. The regime of high planetary mass, far from the star in the parameter space considered by \citealt{sta08} can be compared to the regime of low planetary mass, close to the star in the parameter space considered in the present work, while images at short wavelengths -- tracing the small particles in the present work -- lead to comparable structures. This can be seen from comparison of Fig.~6 (lower left panel) or Fig.~10 of \citet{sta08} to Fig.~\ref{resonances_fo} (model~Ib at thermal re-emission wavelengths, model Id at $24\um$ and at $70\um$) and Fig.~\ref{107146_fo} (models~IIa and~IIb at wavelengths $\leq 70\um$) of the present work.

\citet{wya06} modeled planet-disk interaction in debris disks using dynamical simulations including a migrating planet. Although our approach is substantially different, both works attempt to model structures in debris disks. Therefore, we briefly compare the two approaches. In contrast to \citet{wya06}, we use no migrating planet to initially trap the planetesimals into resonance, but instead assume the dust to be produced through collisions in a ringlike, featureless disk. Furthermore, we use a much simpler approach to mimic dust creation in the planetesimal belt. On the other hand, \citet{wya06} investigated the liberation of particles from the initial resonances only, while we include the possibility that particles are trapped into other resonances than those in which they are initially (they do not have to be in resonance initially at all). Thus, the migrating planet is not necessary to produce resonant structures with our approach.

The differences in the two approaches basically result in an inverse situation in the resulting model images. In \citet{wya06}, very prominent structures are seen at long wavelengths (tracing large particles that remain in the resonances which they are in initially). At short wavelengths, the disks appear smoother, because the particles emitting efficiently at these wavelengths cannot be held in the initial resonance. In contrast, our approach results in a much smoother appearance of the disk at long wavelengths that trace particles that cannot be trapped into resonance efficiently, because they are not moving radially through the system. Hence, strong structures at \mbox{(sub-)mm} wavelengths are only found in our work if the initial dust distribution is placed close to a strong resonance with the planet, or if the dust is close enough to the star that even millimeter-sized particles are significantly affected by Poynting-Robertson drag. On the other hand, small particles are easily trapped into resonances. This is because they move radially through the system due to Poynting-Robertson drag and, thus, can ``find'' those resonances.

\section{Results from the dynamical modeling}
\label{dyn_res}

\begin{figure*}
\centering
\includegraphics[width=1\linewidth]{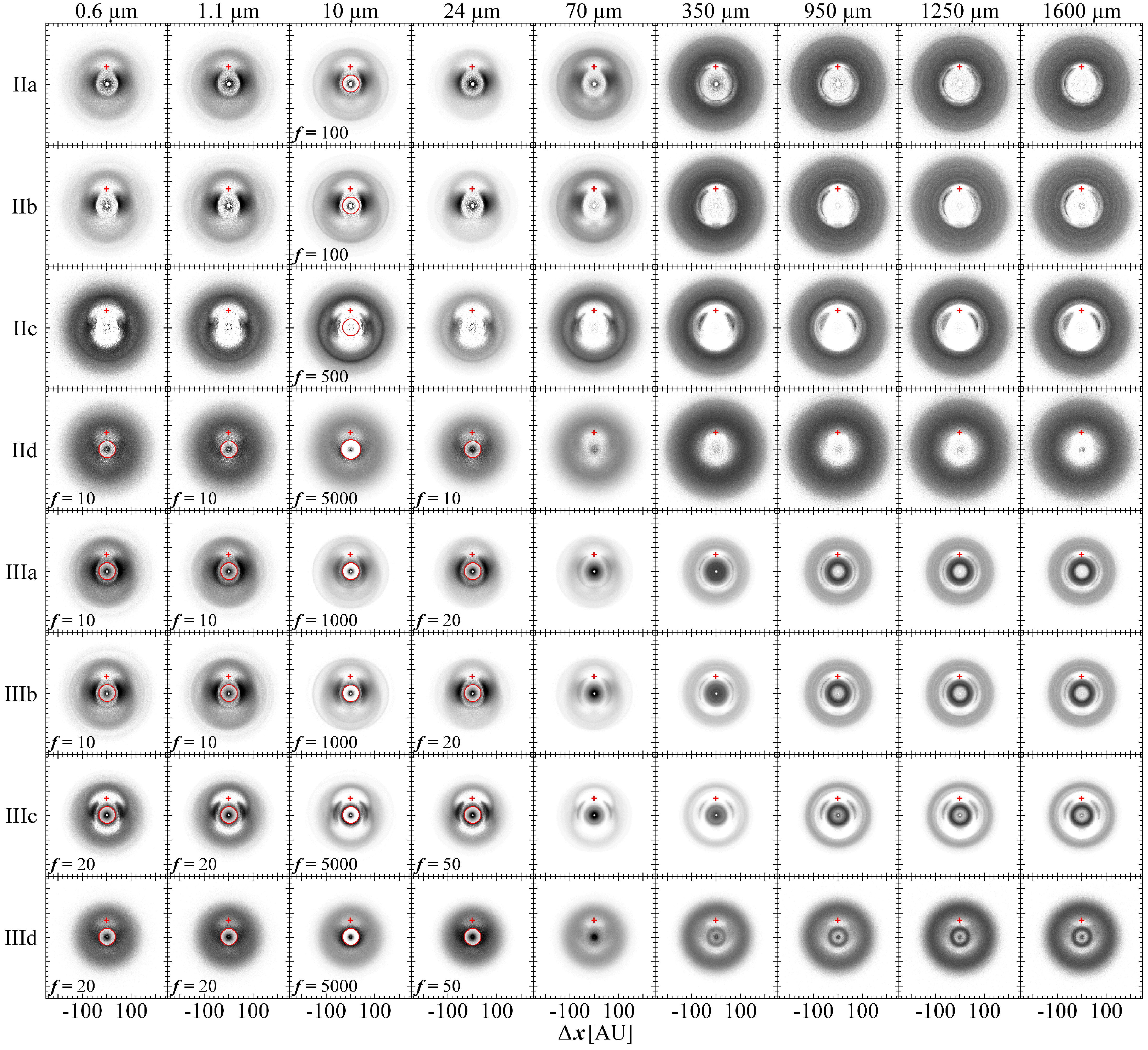}
\caption{Same as Fig.~\ref{resonances_fo}, but for model sequences II and III. The flux is given in arbitrary units and displayed in a logarithmic stretch from zero to $1/2$ of the peak value at wavelengths up to $24\um$ and to twice the peak value at longer wavelengths. Note that a disk without a planet would show no azimuthal structures, but a smooth, featureless radial surface brightness distribution (besides inner and outer disk radius).}
\label{107146_fo}%
\end{figure*}

\begin{figure*}
\centering
\includegraphics[width=1\linewidth]{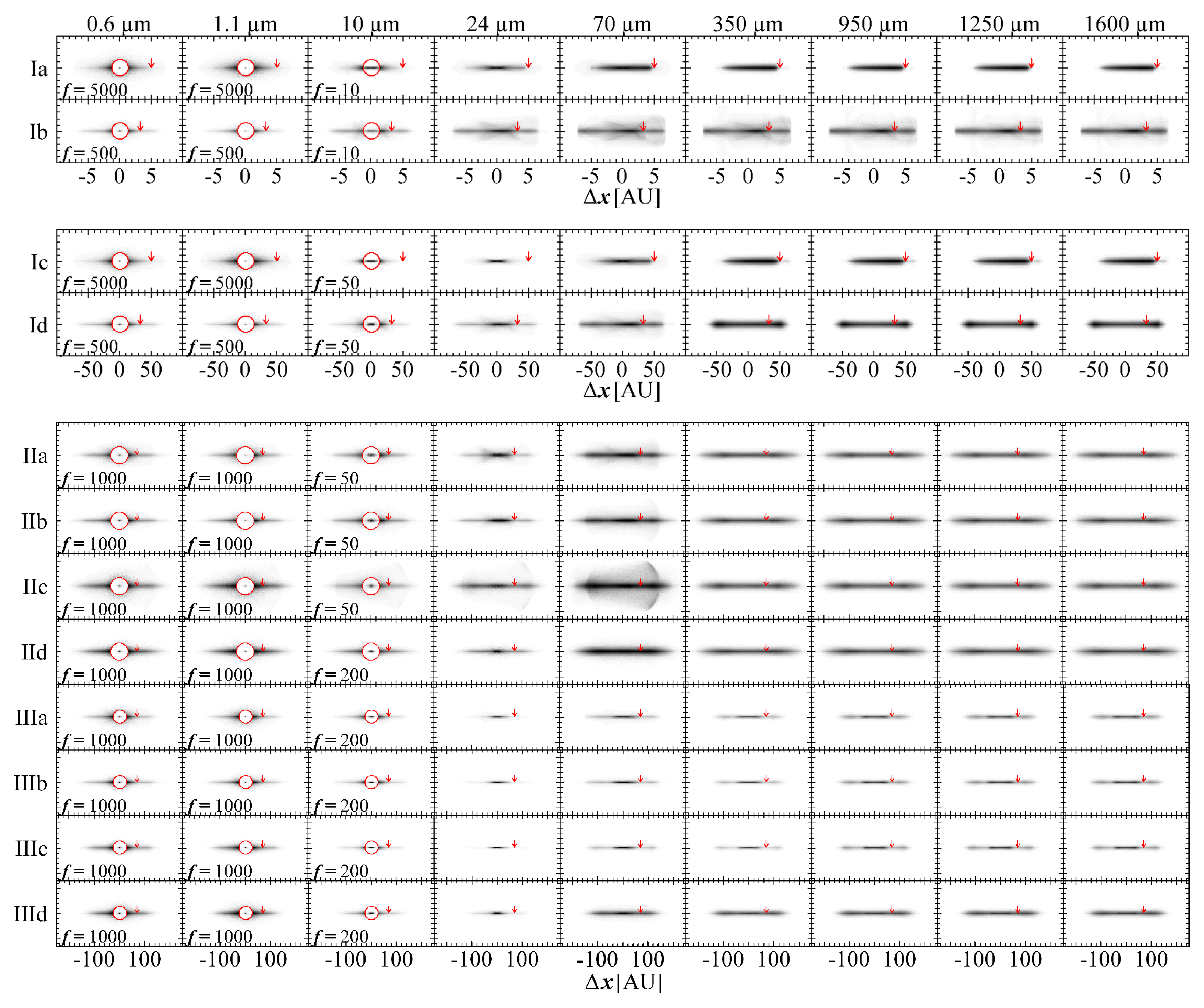}
\caption{Same as Figs.~\ref{resonances_fo} and~\ref{107146_fo}, but for an edge-on orientation of the disk. The position of the planet is indicated by the red arrow in each image. The flux is given in arbitrary units and displayed in a logarithmic stretch from zero to peak value for model sequence~I, from zero to $1/2$ of the peak value at wavelengths up to $24\um$ for model sequence~II and~III, and at longer wavelengths form zero to $1/2$ of the peak value for model sequence II and to the peak value for model sequence III. Note that a disk without a planet would show a smooth, featureless radial surface brightness distribution (besides the possible signpost of inner and outer disk radius).}
\label{images_all_eo}%
\end{figure*}

From the computed images, one can draw several conclusions on the nature and strength of the structures, on the requirements for the generation of strong structures, and on the wavelength dependence of structures in spatially resolved images. The results can be seen in Figs.~\ref{resonances_fo}~to~\ref{images_all_eo} and are briefly discussed in the following.

\subsection{The face-on case}

For a face-on or nearly face-on oriented disk, the azimuthal and radial disk structure can be observed directly. This allows one to draw strong conclusions on the disk structure from observations directly, without the need of detailed modeling or disentangling the real structure from projection effects. On the other hand, it is impossible to draw any strong conclusion on the vertical structure of the disk.

\smallskip
\noindent\emph{Prominent structures at short wavelengths}\\
Observations at short wavelengths (optical and near-infrared scattered light as well as short wavelength thermal re-emission) predominantly trace small grains. These grains are subject to efficient Poynting-Robertson drag which causes them to move radially through the system. Thus, they can easily reach regions in the disk, where they can be trapped into resonance by the planet. This causes prominent structures in the distribution of small grains. The shape and strength of these structures depend on the mass of the planet. In general, a more massive planet is able to keep a larger region clear of dust than a less massive one. On the other hand, less massive planets produce a very prominent, bar-like structure.

\smallskip
\noindent\emph{Prominent structures at small distance from the star}\\
In a similar way as before, particles at small radii are more efficiently affected by Poynting-Robertson drag than particles at larger distance from the star. Thus, the same resonant configuration of planet and disk results in more prominent structures in the distribution of larger grains (at long thermal re-emission wavelengths) if placed closer to the star.

\smallskip
\noindent\emph{Prominent 1:1 resonance at long wavelength thermal re-emission}\\
Large particles are mostly traced by observations at long wavelengths, because they emit more efficiently at these wavelengths and, less important, are cooler than smaller grains. These grains are not moving significantly in radial direction through the system because of Poynting-Robertson drag (as long as they are not too close to the star, see previous point). Only their eccentricity causes a small, periodic change of their radial distance from the star in addition to the dynamical perturbation by the planet. Hence, they can only be trapped into resonance at the radial position they are initially placed. This results in prominent structures caused by 1:1 resonance with the planet, while other resonances can be neglected in most cases. The orbital velocity of these large grains at the same distance from the star like the planet is nearly the same as that of the planet due to the negligible effect of radiation pressure. This renders the resulting structures more stable and, consequently, even more prominent. In the case of a narrow, ringlike shape of the disk which is initially placed close to the planet, this results in a very prominent horseshoe structure (at long wavelength thermal re-emission). If the planet is close to the inner edge or within a broad disk, this results in a horseshoe-like structure in addition to an otherwise nearly featureless disk. If the planet is placed far away from the disk, so that it cannot influence the disk by 1:1 resonance, the effect of the planet is very small. In any case, the prominence of the structures is increasing with increasing mass of the planet.

\smallskip
\noindent\emph{Prominent gaps}\\
If the planet is placed within a broad disk, it opens a ringlike gap in the disk. This gap is very prominent at long wavelengths, while at short wavelengths strong resonant structures dominate the appearance of the disk. At long wavelengths, a horseshoe structure of large particles trapped into 1:1 resonance is visible (see above). The width of the gap and the prominence of the horseshoe structure are increasing with increasing mass of the planet.

\smallskip
\noindent\emph{Faint inner disks}\\
In all our simulations, we always find a more or less significant amount of predominantly small grains which the planet is unable to prevent from moving toward the star. These particles move inward till they reach their sublimation radius. This inner disk, consisting of small, warm grains, is particularly bright at short wavelengths (thermal re-emission and scattered light).

\smallskip
\noindent\emph{Accumulation of dust close to the planet}\\
There is a small amount of dust accumulated close to the planet. These particles are directly captured by the planet. This results in an increased brightness of the planet (and its dust envelope), which increases the chances for direct detection. This scenario has also been suggested to explain the high brightness of Fomalhaut b \citep{kal08}.

\subsection{The edge-on case}

If a disk is seen edge-on, one is faced with a number of challenges when deriving its radial, azimuthal, and vertical structure. Since debris disks are optically thin, one integrates all flux on the line of sight. This results in strong degeneracies between radial and azimuthal structures. Furthermore, vertical structures cannot be assigned to a particular radial and azimuthal position in the disk, but only to a distance from the star projected onto the sky plane. However, most of the known, spatially resolved debris disks are seen close to edge-on. This is most likely because of the higher surface brightness due to more emitting material on the line of sight which, results in an observational bias. For that reason, it is particularly interesting to search for prominent structures in edge-on seen debris disks due to planet-disk interaction, which can be unambiguously identified.

\smallskip
\noindent\emph{Particles scattered out of the disk midplane}\\
Particularly for massive planets close to the radial position of the dust production, there is a halo of particles scattered out of the disk midplane by the interaction with the planet. Therefore, the halo consists mostly of small and intermediate-sized grains moving efficiently radially through the disk. At short to intermediate thermal re-emission wavelengths, resonant structures are visible. These structures are particularly prominent at these wavelengths, because they mostly consist of intermediate-sized grains. These grains are sufficiently affected by Poynting-Robertson drag and radiation pressure to interact with the planet and, thus, to be scattered out of the disk midplane. On the other hand, Poynting-Robertson drag is not strong enough to prevent these grains from being trapped into resonance.

\smallskip
\noindent\emph{Asymmetries in the disk radial brightness profile}\\
There is a brightness asymmetry between the two ansae of the disk. The disk ansa containing the planet is slightly brighter than the opposite ansa. This is due to the accumulation of dust in the resonances which results in clumps of higher density (see face-on case).

\smallskip
\noindent\emph{Multiple peaks and dips in the disk radial brightness profile}\\
A multi-ring structure as seen in particular in the long wavelength results from sequence~III results in a wavy radial brightness profile along the disk midplane in the edge-on case, while peaks are seen at the position of rings, and dips are seen at the position of the gaps in the disk. The strength of these wavy structures depends on the exact configuration of the system and on the observing wavelength.

\subsection{Age dependence of the phenomena}

The dominating effect in our model that causes the dust grains to change their orbits beside the gravitational interaction with the planet is Poynting-Robertson drag. Therefore, one has to compare the Poynting-Robertson time scale of different grains with the age of the systems. This time scale (the time it takes for a particle to spiral onto the star starting at a circular orbit at distance $R_0$) can be approximated following \citet{gus94}:
\begin{equation}
  t_{\rm PR} \approx \frac{400}{\beta} \left(\frac{M_\star}{\rm M_\odot}\right)^{-1} \left(\frac{R_0}{\rm AU}\right)^2 {\rm yr}.
\end{equation}
It is short for small grains (large $\beta$) and increases with grain size (decreasing $\beta$). It also depends on the radial distance from the star. Accordingly, systems with the dust placed closer to the star evolve faster. Furthermore, structures in small grains evolve faster than large grains. The older a system is, the more time there is for larger grains to move significantly in the radial direction, and thereby to be trapped into different resonances. With increasing age of the system, larger grains will have the time to be trapped in prominent resonant structures instead of in the single 1:1 resonant horseshoe structure (in the face-on case). Because these larger grains are traced at longer wavelengths, the transition between the two types of structures will occur at longer wavelengths for older systems.

It is important to note that no collisions between the dust particles are considered in our dynamical simulations. Structures that appear only after a long time or that are particularly traced by particles that are very abundant may be destroyed by chaotic events like collisions. Furthermore, particles may be destroyed through collisions before they are able to form these structures. The strength of this effect is decreasing with decreasing dust mass. Thus, such structures are a result of our modeling, but their significance might be over estimated as compared to the strength of these structures in real systems (particularly in systems with very massive disks).

\subsection{Potential for observations and modeling}

From the above discussions, we found that planet-disk interaction in debris disks produces structures in the disk that allow one to constrain the parameters of a planet-disk system. However, data at one of the wavelength regimes considered are expected to provide only weak constraints on the actual configuration of the system (e.g., planetary mass, major axis, eccentricity, radial distribution of the parent bodies), especially when taking into account observational effects such as noise, resolution effects, and stellar point spread function (PSF) subtraction uncertainties (see Sects.~\ref{obs_ALMA} to~\ref{obs_JWST} for simulated observations and discussion). In the model images themselves, this is particularly obvious, when comparing the images at mid-infrared wavelengths (and shorter) in face-on orientation produced from our runs~II and~III (e.g., at $24\um$) and the corresponding \mbox{(sub-)mm} images (Fig.~\ref{107146_fo}). While the mid-infrared data allow one to distinguish particularly well between planetary mass and eccentricity (e.g., between run~IIa and~IIc), the \mbox{(sub-)mm} data allow one to constrain in particular the major axis of the planet and the position of the parent bodies (e.g., compare images from runs~IIa and~IIIa). Thus, combined observations at both wavelength regimes will be most useful for a detailed analysis of planet-disk systems.

\begin{table*}
\caption{Longest baseline $B_{\rm max}$, shortest baseline $B_{\rm min}$, and resolution characteristics of the array configurations used for the simulations.}
\label{tab_ALMAarrays}      
\begin{center}
\begin{tabular*}{1.0\linewidth}{c@{\extracolsep{\fill}} cccccccccc}
\hline\hline                 
 & & & \multicolumn{2}{c}{Band 10} & \multicolumn{2}{c}{Band 7} & \multicolumn{2}{c}{Band 6} & \multicolumn{2}{c}{Band 5} \\
  \cline{4-5} \cline{6-7} \cline{8-9} \cline{10-11}
  Array  & $B_{\rm min}$ & $B_{\rm max}$ & $\delta_{\rm min}$ & $\delta_{\rm max}$ & $\delta_{\rm min}$ & $\delta_{\rm max}$ & $\delta_{\rm min}$ & $\delta_{\rm max}$ & $\delta_{\rm min}$ & $\delta_{\rm max}$ \\    
  number & [m] & [m] & [$''$] & [$''$] & [$''$] & [$''$] & [$''$] & [$''$] & [$''$] & [$''$] \\
\hline                        
    01   &  15   &   161 & 0.45  & 4.8 & 1.22  & 13.1 & 1.60  & 17.2 & 2.05  & 22.0  \\  
    03   &  15   &   260 & 0.28  & 4.8 & 0.75  & 13.1 & 0.99  & 17.2 & 1.27  & 22.0  \\
    05   &  15   &   390 & 0.19  & 4.8 & 0.50  & 13.1 & 0.66  & 17.2 & 0.85  & 22.0  \\
    07   &  15   &   538 & 0.13  & 4.8 & 0.36  & 13.1 & 0.48  & 17.2 & 0.61  & 22.0  \\
    09   &  15   &   703 & 0.10  & 4.8 & 0.28  & 13.1 & 0.37  & 17.2 & 0.47  & 22.0  \\
    11   &  15   &  1038 & 0.07  & 4.8 & 0.19  & 13.1 & 0.25  & 17.2 & 0.32  & 22.0  \\
    13   &  24   &  1440 & 0.05  & 3.0 & 0.14  &  8.2 & 0.18  & 10.7 & 0.23  & 13.8  \\
    15   &  24   &  1811 & 0.040 & 3.0 & 0.108 &  8.2 & 0.140 & 10.7 & 0.180 & 13.8  \\
    17   &  49   &  2297 & 0.031 & 1.5 & 0.085 &  4.0 & 0.112 &  5.3 & 0.144 &  6.7  \\
    19   &  49   &  3105 & 0.023 & 1.5 & 0.063 &  4.0 & 0.083 &  5.3 & 0.106 &  6.7  \\
    21   &  79   &  6068 & 0.012 & 0.9 & 0.032 &  2.5 & 0.042 &  3.3 & 0.054 &  4.2  \\
    23   &  79   & 11457 & 0.006 & 0.9 & 0.017 &  2.5 & 0.022 &  3.3 & 0.029 &  4.2  \\
  25, 27 &  79   & 14444 & 0.005 & 0.9 & 0.014 &  2.5 & 0.018 &  3.3 & 0.023 &  4.2  \\
\hline                                   
\end{tabular*}
\end{center}
\end{table*}

\section{Evaluating the observability (1): ALMA}
\label{obs_ALMA}

In the following sections, we simulate observations of the modeled disk images using different near-future facilities. In addition, we discuss the detectability of the structures with the {\it Hubble} Space Telescope in order to explain why the prominent scattered light structures seen in our simulations have not been detected yet. In this section, we start with ALMA.

\subsection{General sensitivity study for debris disks}
\label{sect_almasens}

ALMA will allow interferometric observations of debris disks in the \mbox{(sub-)mm} regime with unprecedented sensitivity and spatial resolution. Consequently, it is the most promising instrument to study structures in debris disks in this wavelength regime in the near future. ALMA is a very complex instrument, providing the user with multiple options that allow one to optimize the instrument set-up for the planned observations. A detailed study of the performance of each configuration is necessary to optimize the observational outcome and efficiency. This study in the context of spatially resolved observations of debris disk is described in the following.

\subsubsection{Model description}
\label{sect_imaging_almasens}

\begin{figure}
\centering
\includegraphics[width=\linewidth]{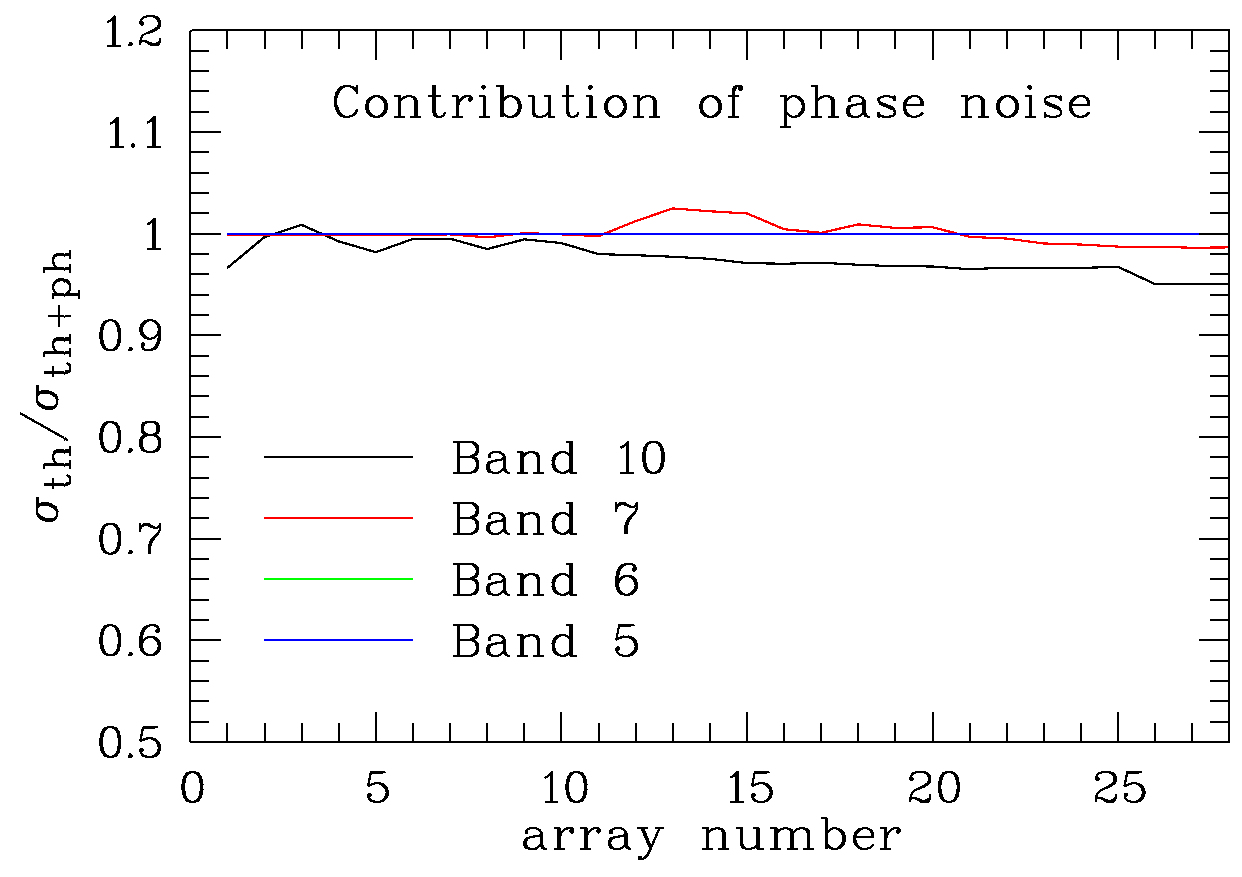}
\caption{Contribution of phase noise to the total noise in our ALMA observations (J. P. Ruge, personal communication). The noise levels are computed as the standard deviation of all pixels in simulated observations of empty images (all pixels set to 0.0). The quantity $\sigma_{\rm th}$ represents the standard deviation in simulations with only thermal noise considered, while $\sigma_{\rm th + ph}$ is derived from simulations considering thermal and phase noise. Ratios higher than 1.0 may occur because of uncertainties in noise estimates and because of different noise contributions that in part cancel out each other. The curves for Band~5 and Band~6 are very close to 1.0 and cannot be distinguished in the plot. In any case, the contribution of phase noise is lower than 5\%.}
\label{fig_phasenoise}%
\end{figure}

All simulations are carried out for a debris disk around a solar-type star (realized as a black body radiator with $R_\star = 1.0\,{\rm R_\odot}$, $L_\star = 1.0\,{\rm L_\odot}$, $T_{\rm eff} = 5778\,{\rm K}$). A simple, analytical disk model is used to keep the results as general as possible. It consists of a circular ring ($r_{\rm out} = 1.1\,r_{\rm in}$) of dust with constant surface density (similar to the width of a ring of dust particles on orbits with identical major axis and eccentricity of 0.1, if the directions of the major axes of the orbits are distributed randomly within the plane of the disk) and an opening angle of $10\degree$. Astronomical silicate with a bulk density of $2.7\,{\rm g/cm^3}$ \citep{dra84, wei01} is employed for the chemical composition of the dust. The grain size distribution follows a power-law distribution (exponent $-3.5$) with lower cut-off size $a_{\rm min} = 0.45\um$ (i.e., the blow-out size of the system) and upper cut-off size $a_{\rm max} = 2\,{\rm mm}$. Only observing wavelengths up to $1.6\,{\rm mm}$ are considered. Therefore, the emission of larger grains is negligible for a sufficiently steep grain size distribution. The inner disk radius is set to 5\,AU, 50\,AU, and 100\,AU in different runs. This covers a variety of known debris disks, e.g., our solar system, $\epsilon$\,Eri \citep{gre98,bac09}, HD\,107146 \citep{ard04,cor09,ert11}, Fomalhaut \citep{kal05}, and AU\,Mic \citep{aug06}. The dust mass of the disk is set to $10^{-8}\,{\rm M_\odot}$ for simulating the model images and scaled to different masses later (assuming an optically thin debris disk) to evaluate the sensitivity. These masses can be converted into fractional luminosities following the equation
\begin{equation}
 \label{eq_m_l}
 \frac{L_{\rm d}}{L_\star} = 4.5\times10^7 \bigg(\frac{r}{\rm AU}\bigg)^{-2} \frac{M_{\rm dust}}{\rm M_\odot},
\end{equation}
i.e., the dust luminosity for a given dust mass and stellar luminosity scales with $r^{-2}$ and is proportional to the dust mass, where $4.5\times10^7$ is the factor for $r = 1\,{\rm AU}$ given the above model of the system (stellar properties and disk properties). This is because the total luminosity of a given dust species only depends on the stellar spectrum, the total surface of the dust ($\propto$ dust mass), and its absorption efficiency, assuming thermal equilibrium (no information about the wavelength range is carried in which the majority of this luminosity is emitted, i.e., about the dust temperature). The resulting factors are $1.8\times10^6$, $1.8\times10^4$, and $4.5\times10^3$ for the disk models with $r_{\rm in} = 5\,{\rm AU}$, $50\,{\rm AU}$, and $100\,{\rm AU}$ in the present study.

Simulations of images from these disk models are performed for face-on and edge-on orientation of the disk. The pixel resolution of the images in AU can then be computed as $2\,r_{\rm out} / 101$ (the total extent of the disk is chosen to be 101 pixels in one direction) and in arcseconds by dividing the result by the distance of the system in pc. Different distances are assumed to simulate observations (Sect.~\ref{sect_simul_almasens}). If the total extent of the images as described above is smaller than five times the resolution element (FWHM) of the observations (in particular for the models with $r_{\rm in} = 5\,{\rm AU}$ at large distance and small array extent), the background fluctuation cannot be quantified in a reliable way. Thus, the resolution of the images is lowered in these cases and the empty region around the disk image is increased. This is done dynamically, so that the pixel resolution is 1/20 of the FWHM and the total extent of the images is 200 pixel ($10 \times {\rm FWHM}$ of each observation).

\subsubsection{Simulation of observations}
\label{sect_simul_almasens}

\begin{figure*}
\centering
\includegraphics[width=1\linewidth]{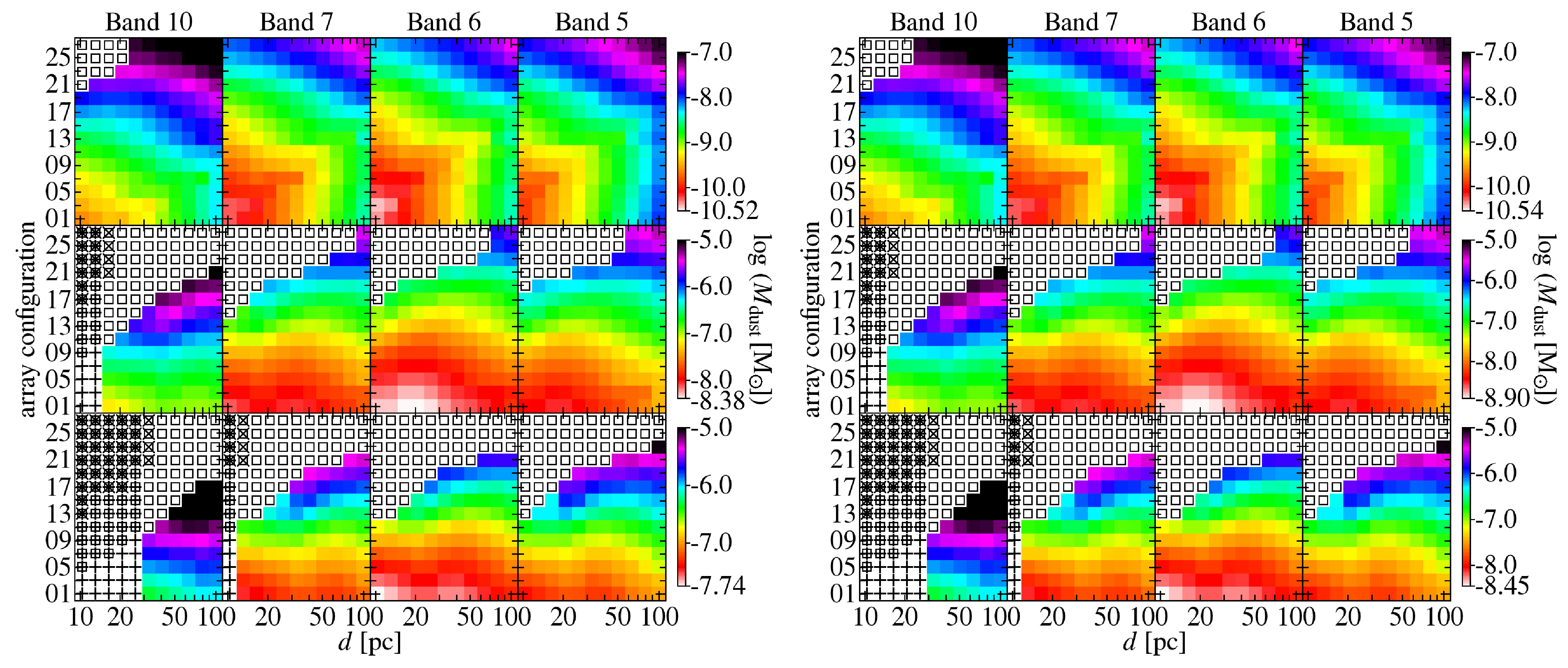}
\caption{Maps of the $10\,\sigma$ sensitivity of ALMA to face-on seen ({\it left}) and edge-on seen ({\it right}) debris disks as a function of the distance and array configuration (spatial resolution) for different wavelengths of observations. The three rows refer to different disk extents given by $r_{\rm in}$ of 5\,AU ({\it top}), 50\,AU ({\it middle}), and 100\,AU ({\it bottom}). Disk mass can be converted into fractional luminosity following Eq.~\ref{eq_m_l}. For details about the model images and about the simulation of observations see Sects.~\ref{sect_imaging_almasens} and~\ref{sect_simul_almasens}. The detectable disk mass is color-coded and can be converted into fractional luminosity following Eq.~\ref{eq_m_l}. White pixels with symbols mark positions where the simulations failed or observations are impossible for various reasons denoted by the symbols as follows (see Sect.~\ref{sect_results_almasens} for more details): {\it cross} -- The spatial extent of the disk is larger than the field of view of a single pointing, {\it x} -- the disk is ``resolved out'' (the radial width of the ring is larger than the largest scale detectable), {\it open square} -- artifacts of the image reconstruction caused by deficiencies of the model image dominate the simulated observations.}
\label{maps_almasens}%
\end{figure*}

The CASA ALMA simulator\footnote{http://casa.nrao.edu/} (procedure {\tt simdata}) is used to simulate observations of the model images. The simulations are conducted using the configurations of the ALMA array in full operations provided by the CASA simulator (using only every second configuration from configurations 01 to 27)\footnote{For a detailed explanation of all parameters used in this procedure, see the corresponding manuals at http://casaguides.nrao.edu/index.php?title=Simulating\_Observa-tions\_in\_CASA}. These array configurations contain 50 12-m antennas each. The baseline ranges of the array configurations are listed in Table~\ref{tab_ALMAarrays}. Bands 10, 7, 6, and 5 (central wavelengths of $350\um$, $950\,um$, $1250\um$, and $1600\um$)\footnote{http://almascience.eso.org/document-and-tools} are used. The exact central wavelength of each bandpass is chosen to avoid strong atmospheric absorption bands. The band width used is 7.5\,GHz for all simulations following the ALMA Cycle 0 Technical handbook\lastfootnote. The objects are placed at an optimal position in the sky reaching the zenith during observations (${\rm RA} = 18^{\rm h}$, ${\rm DEC} = -23\degree$). Total observing time is 8\,h, while the single integration time is 60\,s. Thermal noise is added to the simulated visibilities. Good, but realistic weather conditions are assumed (precipitable water vapor~=~0.65\,mm, ground temperature~=~269\,K\footnote{www.eso.org/gen-fac/pubs/astclim/}). From a comparison with thermal noise, we find phase noise to be negligible (Fig.~\ref{fig_phasenoise}) and, therefore, neglect this contribution to the total noise. For image reconstruction, natural weighting of the visibilities is applied. Deconvolution is performed with 500 iterations and a threshold of 0.01 mJy/beam. Single pointing observations are simulated rather than mosaicking, because most of the simulated images fit into one field of view of a single pointing observation. Mosaicking would require significantly more observing time, increasing the total time on target to an unrealistic amount considering the high pressure on ALMA expected.

For the simulated observations of the model images, the disks are placed at 11 different distances distributed logarithmically from 10\,pc to 100\,pc. The dust mass \mbox{($\propto$ total} disk flux assuming an optically thin disk) is scaled to different values. The S/N is estimated as the ratio between peak flux in the simulated observations of a target and the rms of the background fluctuation. A linear fit is performed on the distribution of S/N over disk mass considering only data with ${\rm S/N} > 20$. From this fit, the disk mass needed to reach an S/N of 10 is computed. This is considered to be a robust detection of the disk.

\subsubsection{Results}
\label{sect_results_almasens}

To properly interpret the results, one has to consider the following specifics of an interferometric observation:
\begin{itemize}
 \item The field of view of an interferometric observation using one single pointing is limited by the FWHM of an observation with one antenna (the primary beam). For the 12-m antennas of ALMA, this is $\lambda / 12\,{\rm m}$.
 \item The spatial resolution $\delta_{\rm min}$ of an observation (the FWHM of the synthesized beam) is determined by the maximum baseline $B_{\rm max}$ following $\delta_{\rm min} = \lambda / B_{\rm max}$.
 \item The largest angular scale $\delta_{\rm max}$ of a target that will not be filtered out (resolved out) by the interferometric observations is determined by the minimum baseline $B_{\rm min}$ following $\delta_{\rm max} \approx 0.6 \lambda / B_{\rm min}$\footnote{ALMA Cycle 0 Proposers Guide, http://almascience.eso.org/ document-and-tools}.
\end{itemize}

The derived values of $B_{\rm min}$ and $B_{\rm max}$ as well as the resolution characteristics $\delta_{\rm min}$ and $\delta_{\rm max}$ for the different array configurations and bands considered are listed in Table~\ref{tab_ALMAarrays}. With the derived values one can now explain the features seen in the sensitivity maps that are shown in Fig.~\ref{maps_almasens}. First, there are a number of configurations in the explored parameter space that cannot be observed with the strategy assumed or where the approach used to estimate the sensitivity produces erroneous results:
\begin{itemize}
 \item In particular very extended, nearby disks, cannot be observed at short wavelengths without mosaicking, since the spatial extent of the disks is larger than the field of view of the observations. The configurations where this is the case are marked by a black cross in an empty (white) pixel in Fig.~\ref{maps_almasens}. Note that parts of the disk can still be observed with one single pointing \citep{bol12}.
 \item For the highest resolution (most extended arrays), in particular nearby, extended disks are resolved out. These regions are marked in Fig.~\ref{maps_almasens} by an ``x''.
 \item For high spatial resolution, the FWHM of the synthesized beam is similar to, or smaller than, the pixel size of the synthetic images. This results in heavy residuals from image reconstruction and does not allow one to properly quantify the S/N in the simulated maps. Note that this is a limitation of the approach, not a real limitation of ALMA. However, at such high resolution, the disks would have to be extremely massive (bright) to be detected at all (see the discussion of the sensitivity below).
\end{itemize}

In the regions of the parameter space at which the simulations are not affected by the above limitations, one can evaluate the sensitivity. The following behavior in the parameter space can be found:
\begin{itemize}
 \item In general, the highest sensitivity to disk mass is reached in Band 6 (central wavelength $1250\um$).
 \item For a given spatial resolution, the sensitivity is only slightly decreasing with increasing distance as long as the disk is spatially resolved (${\rm FWHM} \leq 2\,r_{\rm out}$). This can be explained by the fact that the surface brightness of the disk (e.g., in mJy/AU$^2$) is decreasing with the distance squared, but the beam covers an increasing area of the disk, depending on the scales of the dominating disk structures.
 \item For a given distance, the S/N is increasing with decreasing spatial resolution as long as the disk is spatially resolved. This is because the beam covers a larger area of the disk at very similar sensitivity (e.g., in mJy/beam). As long as the FWHM is smaller than the dominating structure of the disk (e.g., the ring width), the sensitivity is increasing with the square of the FWHM.
 \item As soon as the disk is spatially unresolved, the sensitivity drops with the distance squared for a given spatial resolution.
 \item In general, disks seen edge-on have a higher surface brightness. Thus, spatially resolved observations of an edge-on seen disk will result in a higher S/N than the same observations of the same disk seen face-on. In addition to that, there is no significant difference between imaging of edge-on and face-on disks (for the employed disk models). This can be explained by the fact that for spatially resolved imaging only the peak S/N is considered, which depends on the scale and brightness of the brightest structures in the disk. This is in both cases the radial width of the ring and, for disk seen edge-on, the vertical height (which is of the same order in our model). For spatially unresolved observations, the orientation of the disk has no effect at all.
\end{itemize}
The above qualitative discussion leads to an important recommendation for optimal spatially resolved imaging of debris disks with ALMA. As long as the structures of interest are bright enough, one should always use a spatial resolution similar to the scales of interest. Most importantly, one should avoid to over-resolve the structures, because sensitivity then decreases with the square of the spatial resolution.

ALMA will not only increase our knowledge about the known spatially resolved debris disks owing to the higher sensitivity and spatial resolution. Because of the nearly constant sensitivity to dust mass in spatially resolved imaging at distances up to $\approx 100\,{\rm pc}$ (for disks with radial extents of few tens of AU), ALMA will significantly increase the sample of debris disks spatially resolved at millimeter wavelengths toward more distant systems compared to earlier instruments.

\subsection{Observability of planet-disk interaction}

From our dynamical modeling, we select several runs that result in prominent structures at wavelengths $\geq 350\um$ in face-on and/or edge-on orientation. The total flux in the selected images is scaled to the total flux of selected, well-studied debris disks (Table~\ref{reference}). The CASA simulator is used to predict results of observations. The optimum array configuration is selected by the resulting S/N and spatial resolution in the simulated images.

\subsubsection{Selection of runs and preparation of the images}

\begin{table*}
\caption{Reference systems for our simulated ALMA observations}
\label{reference}
\begin{center}
\begin{tabular}{cccccccccc}
\hline\hline                 
 Run & Reference & $\lambda_{\rm ref}$ & $d$ & Host star & Age & $R_{\rm dust}$ & $F_{\rm disk}$ at $\lambda_{\rm ref}$ & $M_{\rm disk}$ & Ref. \\
      &                 & [$\um$] & [pc] &       & [Myr] & [AU]      & [mJy] & [$\rm M_\odot$]   &      \\
\hline
 Ib              & $\epsilon$\,Eri & 24      & 3.2  & K2\,V & 850   & 3.0       & 330.0 & $5.7\times10^{-11}$ & 1    \\
 Ic              & HD\,105         & 850     & 40.0 & G0\,V &  30   & 45 -- 120 & 10.7  & $1.2\times10^{-7}$  & 2    \\
 IIc             & HD\,107146      & 350     & 29.5 & G2\,V & 130   & 50 -- 250 & 319.0 & $7.3\times10^{-7}$  & 3, 4 \\
 IIIa            & HD\,107146      & 350     & 29.5 & G2\,V & 130   & 50 -- 250 & 319.0 & $3.7\times10^{-7}$  & 3, 4 \\
 IIId            & HD\,107146      & 350     & 29.5 & G2\,V & 130   & 50 -- 250 & 319.0 & $5.5\times10^{-7}$  & 3, 4 \\
\hline
\end{tabular}
\tablebib{(1)~\citet{bac09}; (2)~\citet{nil10}; (3)~\citet{cor09}; (4)~\citet{ert11}.}
\end{center}
\end{table*}

\begin{figure*}
\centering
\includegraphics[width=\linewidth]{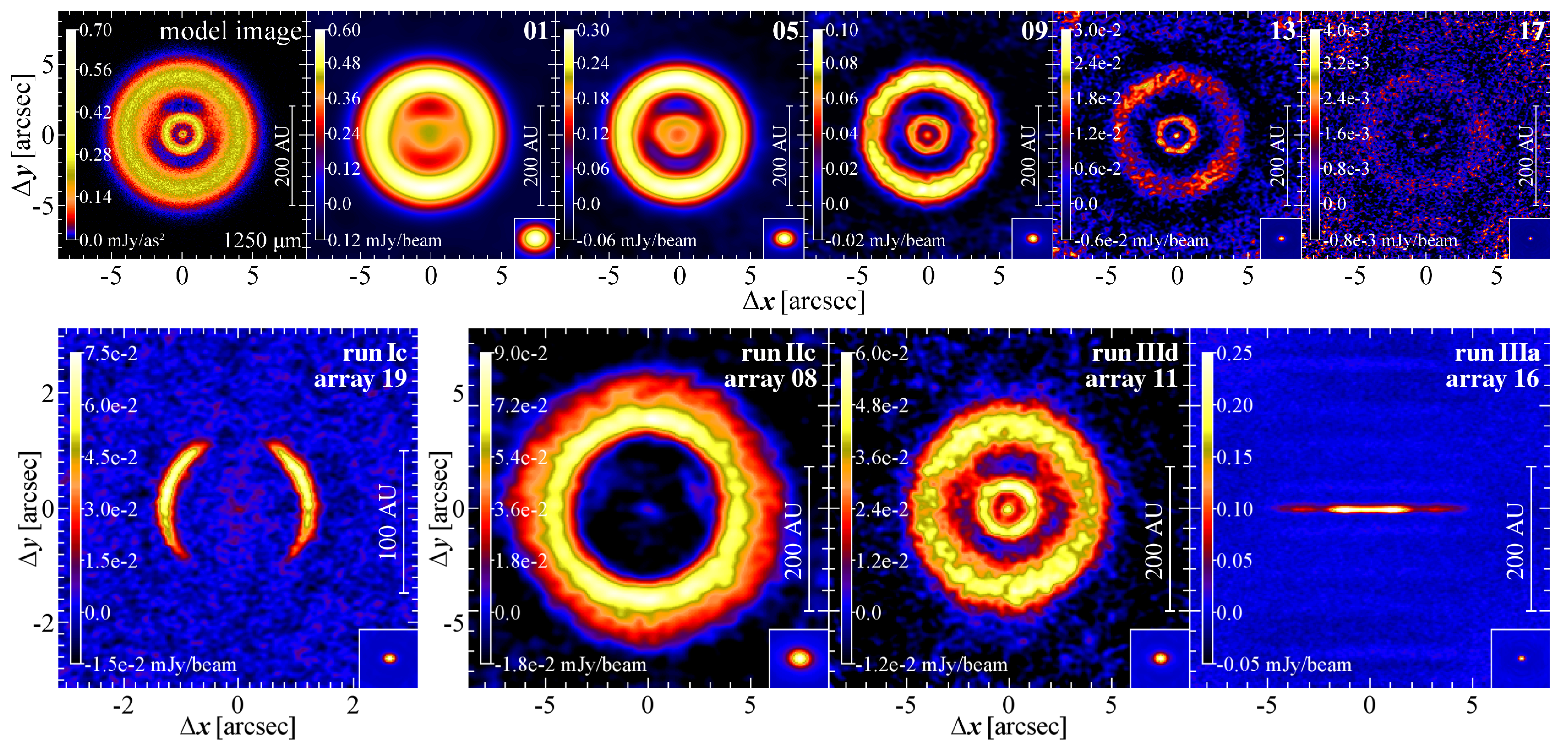}
\caption{\emph{Top:} Simulated ALMA observations of the model image resulting from run IIId (face-on orientation) for different array configurations. Only the results from a representative number of array configurations are shown. The number in the upper-right corner of each image denotes the array number. The corresponding beam is displayed in the lower-right corner of each image. \emph{Bottom:} Simulated ALMA observations of all selected model images. The optimum array configurations have been selected. The run and the array number are displayed in the upper-right corner, the corresponding beam is displayed in the lower-right corner of each image. All observations were simulated for Band~6 (central wavelength $1250\,\um$).}
\label{sequ_alma}%
\end{figure*}

The runs for which ALMA simulations are carried out are selected according to the prominence of the structures seen in the \mbox{(sub-)mm}:
\begin{itemize}
 \item Runs~Ib, Ic, IIc, and~IIId are found to represent the prominent structures in the synthetic images well for face-on orientation of the disk (Fig.~\ref{resonances_fo} and~\ref{107146_fo}). 
 \item Run~IIIa is selected as an example for structures seen in edge-on orientation of the disk (Fig.~\ref{images_all_eo}).
\end{itemize} 
From these runs, observations are simulated at Band~6 (central wavelength $1250\um$, found in Chapt.~\ref{sect_almasens} to be the most sensitive one for the intended observations) as follows:
\begin{itemize}
 \item Known debris disks with radial dust distributions similar to the simulated ones are selected as reference objects. Systems around solar-type stars are considered, because solar-type stars are considered for the dynamical simulations as well.
 \item A representative photometric measurement (in the \mbox{(sub-)mm}, where possible) is used to scale the total flux of the simulated images. Therefore, the total flux of the disk at the wavelength of these observations is computed.
 \item The disk is placed at the distance of the reference debris disk. A flux-scaling factor is derived by comparing the flux of the modeled disk with the measured flux at the reference wavelength. This scaling factor is applied to the image at Band~6 to properly scale the flux in this image to a realistic value.
 \item  The stellar contribution of the host star of the reference debris disk is added to the center of the synthetic image.
\end{itemize}
The reference debris disk for each run, the reference wavelength, the applied flux at this wavelength, some basic information about the reference system, and the resulting dust mass when scaling the model disk to the observed flux are shown in Table~\ref{reference}. It is important to note that the model scaled to the observed flux of the reference debris disk at one wavelength is not expected to reproduce the whole spectral energy distribution of the disk properly. However, the reference debris disks are selected by their similarity to the model systems in the radial dust distribution and Band~6 is clearly in the Rayleigh-Jeans regime of the dust re-emission. Accordingly, the flux at this wavelength is considered to be at least realistic for a debris disk similar to the reference disk.

The reference debris disks used for the scaling of the model are motivated as follows:
\begin{itemize}
 \item $\epsilon$\,Eri for run Ib: The inner debris ring \citep[$\sim 3\,{\rm AU}$ from the star;][]{bac09} is used. For this ring, the $24\um$ flux is used as reference value to scale the flux in all model images, since the other disk components are expected to have significant contribution to the flux at longer wavelengths.
 \item HD\,105 for run Ic: HD\,105 is a massive debris disk around a solar-type star. Most of the dust is expected to be concentrated at a distance of 40\,AU to 50\,AU from the star \citep{hil08}.
 \item HD\,107146 for runs IIc, IIIa, and IIId: The work on the HD\,107146 disk \citep{ert11} was the motivation for these runs (Sect.~\ref{sect_initial}).
\end{itemize}

\subsubsection{Parameters for the CASA simulations}

The simulations are carried out using all configurations of the ALMA array in full operations provided by the CASA simulator. Observing conditions and parameters for the simulations are identical to the simulations in Sect.~\ref{sect_almasens}. Band~6 (central wavelength $1250\um$) is used. Single pointing observations are simulated rather than mosaicking, since all simulated images fit into one field of view of a single pointing at Band~6.

\subsubsection{Results}

We find from the simulations of observations at different spatial resolution (different maximum baseline) that a trade-off between sensitivity and spatial resolution is necessary (see also Sect.~\ref{sect_results_almasens}). This is particularly well illustrated in the example of run~IIId (Fig.~\ref{sequ_alma}, top). The bar-like structure within the (outer) gap in the disk is only visible at the lowest resolutions, while one needs a higher resolution to clearly resolve the inner gap and to clearly separate the innermost peak from the inner ring of dust at $1''$ to $2''$ from the star. For the five model disks considered for the ALMA simulations, the best array configurations based on high S/N and high resolution are selected. The model image resulting from \emph{run~Ib} is found to be too faint to be detected by any of the simulated observations. For the other four model images, the results are shown in Fig.~\ref{sequ_alma} (bottom). The selection of the optimum array configuration for each case is described in the following. The results are consistent with the suggestions for optimal spatially resolved imaging of debris disks derived in Sect.~\ref{sect_results_almasens}.

For \emph{run~Ic}, the structures are very smooth and no substructure is seen. Thus, one can use the peak S/N as a good tracer for the significance of the structures detected. Based on an additional, visible inspection of the images, a peak S/N of 22 (array~19, \mbox{FWHM = $0\farcs11 \times 0\farcs15$}) is found to give the best results, while the next larger array would result in a peak S/N of 15, but large parts of the disk would be detected at an ${\rm S/N} < 10$.

For \emph{run~IIc}, the structure of interest would be the faint horseshoe structure at the inner edge of the disk. This structure can in any case only be observed at an S/N of up to 4. This is reached when the size of the beam is comparable to the size of this structure. The optimum is reached using array~08 \mbox{(FWHM = $0\farcs62 \times 0\farcs78$)}. It is important to note that there is a negative background that is nearly homogeneous over the whole image. This can be attributed to the limitations of ALMA to observe very extended structures and of the simulations as described in Sect.~\ref{sect_results_almasens}. The signal has to be evaluated as the flux above this homogeneous background.

For \emph{run~IIId}, value is placed on the multi-ring structure, while the bar-like structure is ignored. An example for an array configuration that results in a significant detection of the bar-like structure can be seen in Fig.~\ref{sequ_alma} (top). A peak S/N of 10 (array~11, \mbox{FWHM = $0\farcs38 \times 0\farcs47$}) is found to result in significant detection of the three disk components (outer and inner ring as well as innermost accumulation of dust), since they have a very similar surface brightness. The bar-like structure is visible at low S/N in these simulated observations, as expected from the previous discussion about the connection between spatial resolution and sensitivity to surface brightness.

For the disk from \emph{run~IIIa} seen edge-on, the structure of interest is the gap in the disk seen as a dip in the radial brightness distribution and the outer disk seen as a secondary peak of the surface brightness beyond this gap. An S/N of 15 (array~16, \mbox{FWHM = $0\farcs16 \times 0\farcs22$} -- similar to the vertical extent of the disk) in the secondary peak of the surface brightness results in a clear detection of both features and allows one to marginally resolve the vertical extent of the disk.

These examples illustrate the findings in Sect.~\ref{sect_results_almasens} for structured disks. ALMA is not able to reach both high sensitivity to surface brightness and high spatial resolution simultaneously. The highest sensitivity at reasonable spatial resolution is reached if the
resolution of the observations is similar to the scale of the structures expected, as long as these structures are bright enough.

\section{Evaluating the observability (2): The Hubble Space Telescope}
\label{obs_hst}

From the dynamical simulations, prominent structures in debris disks are found, in particular in scattered light and short-wavelength thermal re-emission. In contrast, no debris disk seen face-on is known to exhibit such structures in scattered light observations. Therefore, we briefly examine the observability of such structures in face-on seen disks with present instruments in scattered light, in particular with the HST with which most of the scattered-light detected debris disks have been discovered. Our example here is the HD\,107146 debris disk. It is one of the few debris disks seen (close to) face-on detected in scattered light \citep{ard04, ert11}.

\begin{figure}
\centering
\includegraphics[width=\linewidth]{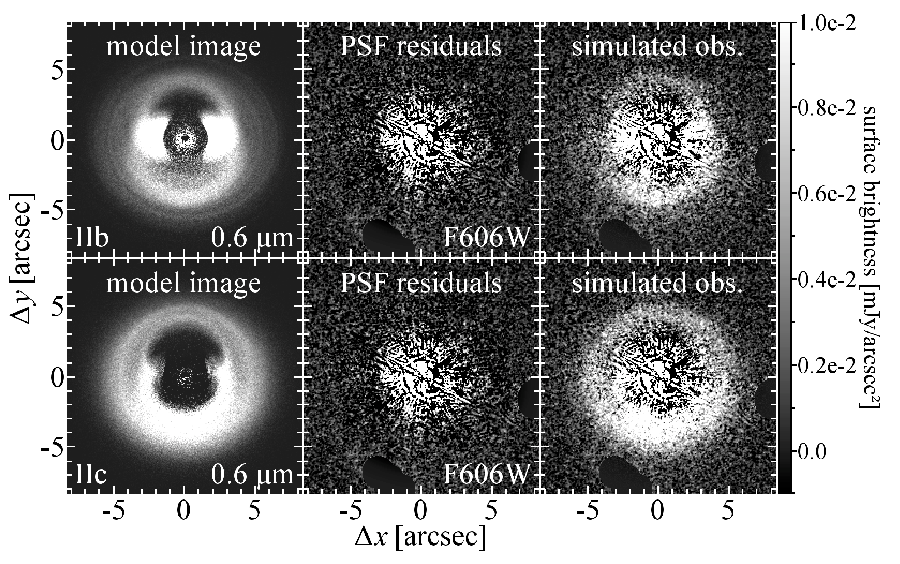}
\caption{Simulated HST/ACS coronagraphic observations of the simulated images from runs IIb and IIc at $0.6\um$. The disk is inclined by $25\degree$ along the $x$-axis, so that the lower half of the disk points toward the observer.}
\label{fig_hst}%
\end{figure}

The HST/ASC image in the F606W ($\lambda_{\rm c} = 0.6\um$) filter presented by \citet{ard04} is used to evaluate the observability of simulated structures in these data. Therefore, the model-subtracted image of \citet{ert11} is employed, which represents the pure PSF subtraction residuals (within the capabilities of the modeling). To these data, simulated images from the dynamical modeling \citep[inclination of $25\degree$ from face-on applied from][]{ard04} are added after scaling them to the scattered-light flux of the fitted disk at the same wavelength. This results in simulated observations of the model images using the same strategy as for the original observations. The results for the model images from runs~IIb and~IIc are shown in Fig.~\ref{fig_hst}.

While the disk is clearly detected, the structures of the modeled images are corrupted by the strong PSF subtraction residuals. Thus, one would not be able to draw strong conclusions on the azimuthal structure of the disk from these data. Furthermore, there is only a small number of debris disks seen face-on and detected in scattered light. In contrast, the two most prominent debris disks seen edge-on, $\beta$\,Pic and AU\,Mic, exhibit clumpy and warped structures that might be interpreted as structures in the disk induced by planet-disk interaction \citep[e.g.,][]{liu04, gol06}. All this suggests that the structures found by the modeling in this work may be present, but not detected in face-on seen debris disks, yet.

\section{Evaluating the observability (3): Space-based near- to mid-infrared telescopes}
\label{obs_JWST}

Near-future space-based near- to mid-infrared telescopes (e.g.,  the {\it James Webb} Space Telescope, JWST, \citealt{gar06}, or the Space Infrared Telescope for Cosmology and Astrophysics, SPICA, \citealt{swi09}) are expected to provide high-sensitivity, high angular resolution imaging capabilities in the optical to near-infrared. In this section, observations with such facilities are predicted based on available information about the expected capabilities of the JWST. Since the focus of this work is on spatially resolved imaging capabilities, the two relevant instruments are the Near InfraRed Camera (NIRCam) and the Mid InfraRed Instrument (MIRI). NIRCam provides imaging capabilities in the $0.6\um$ -- $5.0\um$ wavelength range. In this range, coronagraphy is necessary to block the direct stellar radiation. The HST is very successful in coronagraphic imaging of bright debris disks and the JWST will exceed these capabilities due to its higher sensitivity and spatial resolution. However, the performance of coronagraphic instruments depends on several influences such as PSF stability. Consequently, the exact results of such observations are hard to predict without detailed knowledge about the performance of the instrument during science operations. Hence, the focus of the present section is placed on predictions on the capability of debris disk observations through none-coronagraphic imaging as are possible with MIRI.

\begin{table}
\caption{Reference systems for our simulated JWST observations}
\label{reference_JWST}
\begin{center}
\begin{tabular*}{\linewidth}{c@{\extracolsep{\fill}}ccccc}
\hline\hline                 
 Run             & Reference         & $F_{\rm disk}$ & $F_\star$      & $M_{\rm disk}$    & Ref. \\
                 &                 & [mJy]          & [mJy]          & [$\rm M_\odot$]     &      \\
\hline
 Ia              & $\epsilon$\,Eri & 330.0          & 1726           & $8.5\times10^{-12}$ & 1    \\
 Ib              & $\epsilon$\,Eri & 330.0          & 1726           & $7.3\times10^{-12}$ & 1    \\
 IIa             & HD\,107146      &  19.0          & 40.8           & $4.4\times10^{-8}$  & 2    \\
 IIc             & HD\,107146      &  19.0          & 40.8           & $1.4\times10^{-7}$  & 2    \\
 IId             & HD\,107146      &  19.0          & 40.8           & $5.7\times10^{-8}$  & 2    \\
 IIIc            & HD\,107146      &  19.0          & 40.8           & $7.4\times10^{-9}$  & 2    \\
\hline
\end{tabular*}
\tablefoot{The reference wavelength at which the fluxes are given is $24\um$ for both systems. Also note the information given in Tab.~\ref{reference} about the two reference systems.}
\tablebib{(1)~\citet{bac09}; (2)~\citet{hil08}.}
\end{center}
\end{table}

\begin{figure}
\centering
\includegraphics[width=0.7\linewidth]{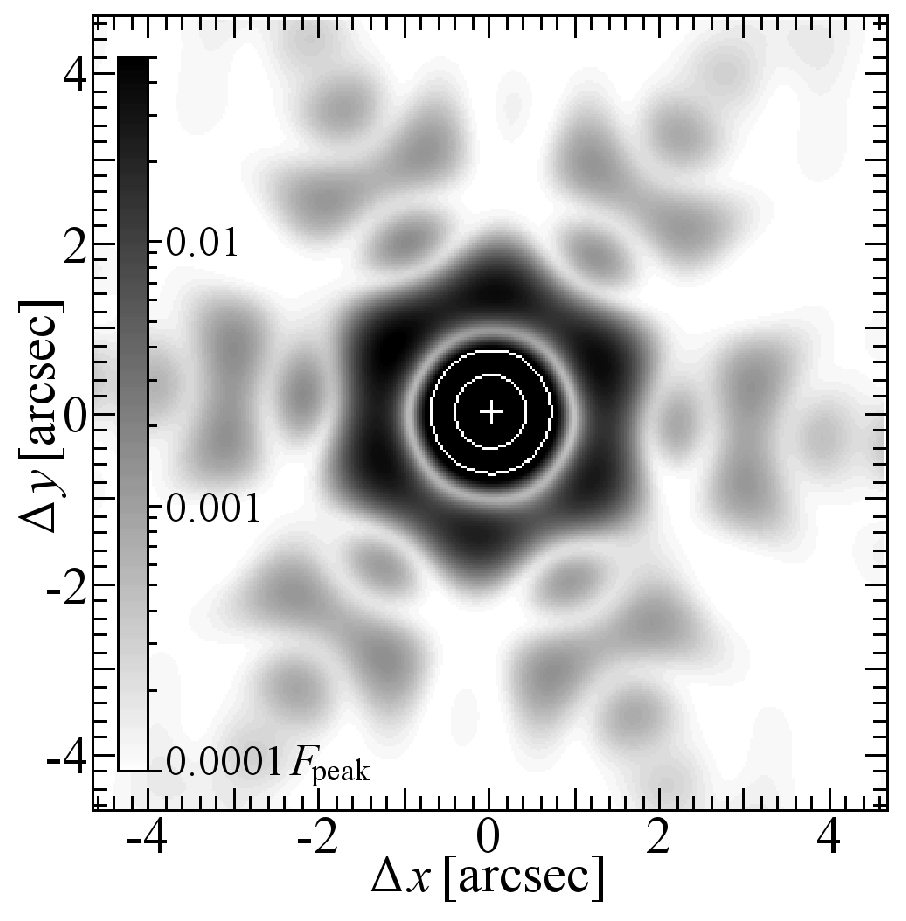}
\caption{Simulated PSF of JWST/MIRI at $25.5\um$. The image is shown in a logarithmic stretch from 0 to 5\% of the peak flux to highlight the high order structure. The white contours represent 10\% and 50\% of the peak flux.}
\label{fig_psf}%
\end{figure}

MIRI will provide imaging capabilities in the wavelength range of $5.0\um$ -- $29\um$. In particular at the long wavelength edge of this range, thermal radiation of debris disks is expected to have a detectable level. With a spatial resolution of $\approx 1''$ and high sensitivity, MIRI is the first instrument that will allow for resolved imaging of a large number of debris disks in this wavelength range.

\subsection{Selection of runs and approach to simulate observations}

Several runs are selected from the dynamical simulations to explore the capabilities of MIRI to detect the structures. A summary of the considered runs and the references used for each run is given in Table~\ref{reference_JWST}. Since $\epsilon$\,Eri is the only known debris disk this close and it is seen close to face-on, only the face-on orientation is used for the simulated observations on the example of the inner disk in this system. For the simulated observations on the example of the HD\,107146 debris disk, face-on and edge-on orientations are used, since a number of disks with different orientations are known at distances of few tens of AU (e.g., AU\,Mic, \citealt{kal04}, q$^1$\,Eri, \citealt{lis10}, HD\,207129, \citealt{kris10}).

\begin{figure*}
\centering
\includegraphics[width=0.9\linewidth]{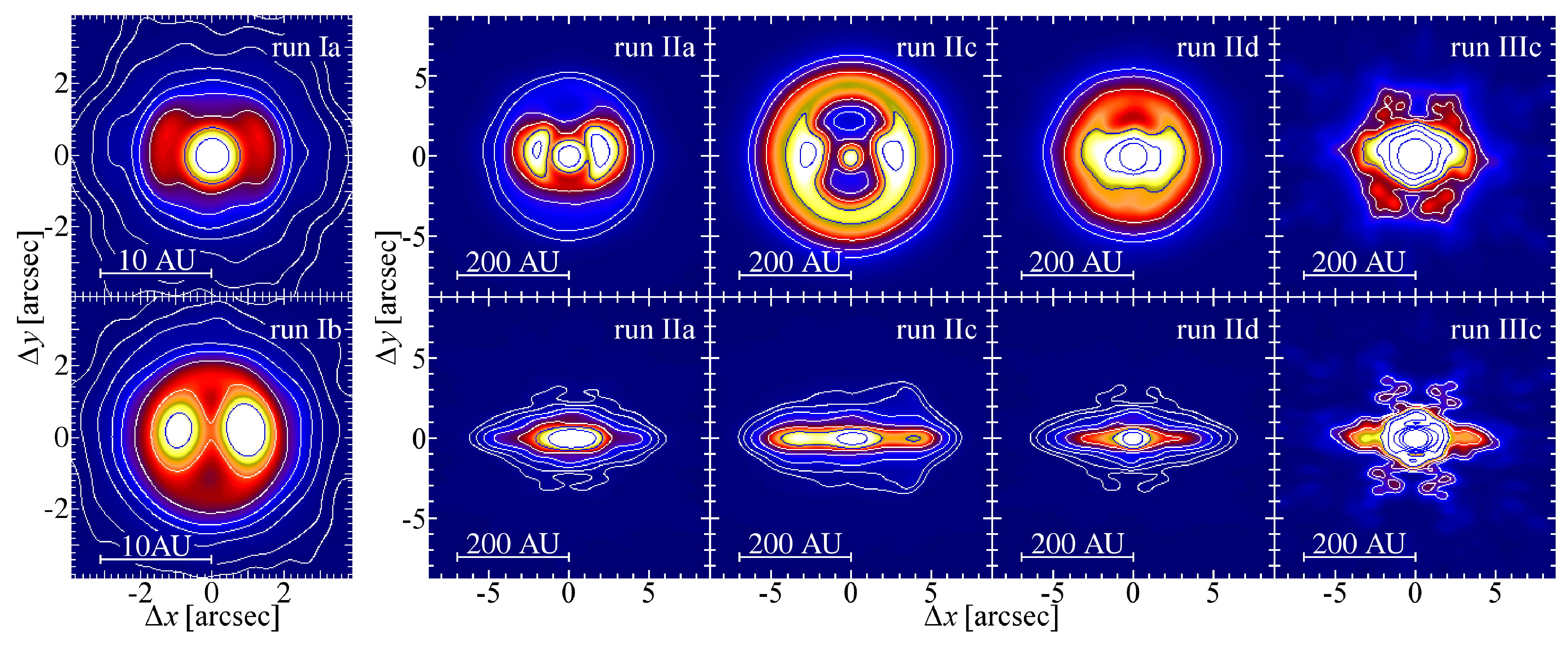}
\caption{JWST/MIRI images at $25.5\um$. \emph{Leftmost column:} Images produced from the runs~Ia and~Ib. The $\epsilon$\,Eri inner debris disk is used as reference. \emph{Columns two to five:} Images produced from the runs~IIa, IIc, IId, and~IIIc. The HD\,107146 debris disk is used as reference. The contours in each image represent $3\,\sigma$, $5\,\sigma$, $10\,\sigma$, $20\,\sigma$, etc.\ (double every step) S/N levels for a total integration time of 10\,s ($\epsilon$\,Eri) and 900\,s (HD\,107146). The images are shown in a logarithmic stretch.}
\label{sequ_jwst}%
\end{figure*}

Because the star contributes significantly to the flux at mid-infrared wavelengths, a stellar PSF subtraction is expected to be necessary to clearly reveal any resolved disk structure. To evaluate the observability of structures in debris disks, a two-step approach is used:
\begin{itemize}
 \item[(1)] Observations of the disk only are simulated by convolving the synthetic disk images with the telescope PSF. From the resulting images, the sensitivity needed is estimated and the capability to spatially resolve the structures is evaluated.
 \item[(2)] The stellar PSF contribution is estimated and the PSF subtraction accuracy needed is evaluated.
\end{itemize}

For \emph{simulated observations of the disk only}, the images at $24\um$ resulting from the dynamical simulations are scaled to the $24\um$ fluxes of two reference debris disks, $\epsilon$\,Eri and HD\,107146. This approach is expected to result in a realistic mid-infrared brightness of the disk images, which is important for the simulated observations, but not in a good reproduction of the SED of the system over the whole wavelength range. The disks are placed at the distance of the reference debris disks. The images are convolved with a simulated PSF of MIRI at $25.5\um$ (Fig.~\ref{fig_psf}) produced by the software {\tt WebbPSF}\lastfootnote. The resulting sensitivity of the observations is estimated using a sensitivity of $30\,\mu{\rm Jy}$ ($10\,\sigma$, $\lambda = 25.5\um$, integration time = $10^4\,{\rm s}$)\footnote{http://www.stsci.edu/jwst/science/data-simulation-resources; effective Aug.~2011} and scaling it with the square root of the actual integration time \citep{gar06}.

\begin{table}
\caption{PSF subtraction accuracy necessary for observations of the modeled debris disks.}
\label{psf_subtraction}
\begin{center}
\begin{tabular*}{\linewidth}{c@{\extracolsep{\fill}}ccccc}
\hline\hline                 
 Run             & $\rho_{\rm disk}$   & $F_{\rm disk}$ & $\rho_{\rm peak}$ & $F_{\rm PSF}$  & $\sigma$   \\
                 & [arcsec] & $\left[\frac{\rm mJy}{\rm beam}\right]$     & [arcsec]          & $\left[\frac{\rm mJy}{\rm beam}\right]$     & [\%]       \\
\hline
 Ia              & 1.0      & 21.8           & 1.4               & 90.0           &  2.4       \\
 Ib              & 1.0      & 55.6           & 1.4               & 90.0           &  6.2       \\
 IIa (fo)        & 2.0      &  1.3           & 1.4               & 2.1            &  6.2       \\
 IIa (eo)        & 2.0      &  1.1           & 1.4               & 2.1            &  5.2       \\
 IIc (fo)        & 2.6      &  0.6           & 2.3               & 0.2            & 30.0       \\
 IIc (eo)        & 3.0      &  1.3           & 3.2               & 0.2            & 65.0       \\
 IId (fo)        & 3.0      &  0.2           & 3.2               & 0.2            & 10.0       \\
 IId (eo)        & 3.0      &  0.9           & 3.2               & 0.2            & 45.0       \\
\hline
\end{tabular*}
\end{center}
\tablefoot{The quantity $\rho_{\rm disk}$ denotes the (projected) distance from the star at which the disk flux is measured (the brightest structures of intent), $\rho_{\rm peak}$ denotes the distance from the star of the closest peak of the stellar PSF\footnotemark, $F_{\rm disk}$ and $F_{\rm PSF}$ are the surface brightness of the disk and the PSF structures at the corresponding distances, and $\sigma$ denotes the accuracy at which the stellar PSF has to be subtracted to reduce the PSF structures to $1/10$ of the disk structure (i.e., $1/10 \times F_{\rm disk}/F_{\rm PSF}$). Values are given for a face-on (fo) and edge-on (eo) orientation of the disk.}
\end{table}

In Fig.~\ref{fig_psf}, one can clearly see the hexagonal structure and the peaks of the PSF at high order. Since the star contributes significantly to the flux in the mid-infrared, one has to perform accurate PSF subtraction to distinguish between real disk structures and these high order PSF structures. To evaluate the \emph{accuracy of stellar PSF subtraction necessary}, the contribution of the PSF depending on the radial distance from the star has to be computed. Therefore, the PSF used to convolve the disk images is scaled to the total flux of the star in each of the two reference systems. The maximum of the PSF in one-pixel wide radial bins (pixel scale of the simulated disk images) is computed. The result is shown in Fig.~\ref{plot_psf}. To obtain a reliable image of the disk, the contribution of the stellar PSF should be reduced by PSF subtraction (e.g., through observations of a reference star) to a level of $\approx 1/10$ or less of the disk flux at a comparable distance from the star.

\subsection{Results}

The results from the \emph{simulated observations of the disk only} are shown in Fig.~\ref{sequ_jwst}. In both the $\epsilon$\,Eri and the HD\,107146 case, one can see that the disks are spatially resolved. One can clearly distinguish between the different planet-disk configurations in both the face-on and the edge-on case. Only the results of run~IIIc do not show a clearly resolved image of the disk. In contrast, the bright, unresolved inner part of the disk dominates the emission, which results in a bright image of the PSF overlayed on the outer disk structure.

\begin{figure}
\centering
\includegraphics[width=\linewidth]{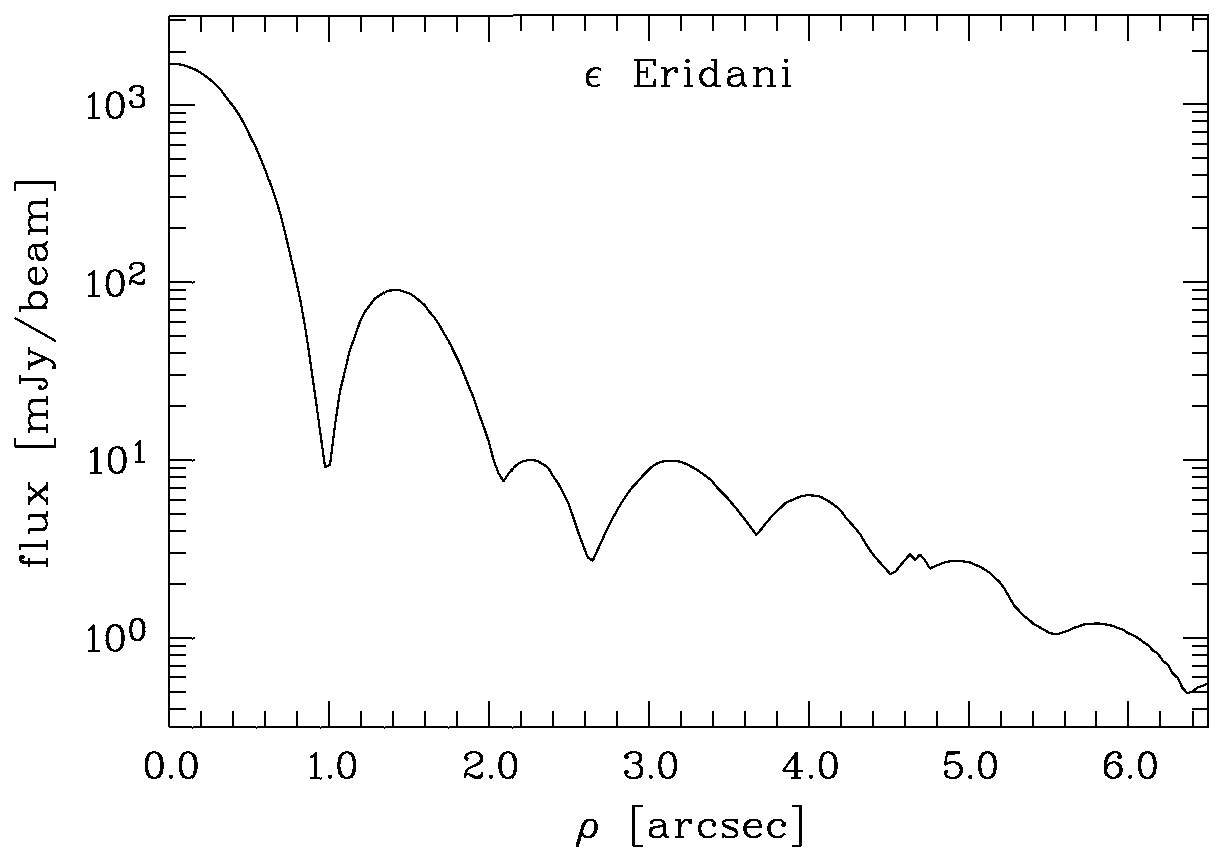}
\caption[Radial distribution of the JWST/MIRI PSF structures at $25.5\um$.]{Radial distribution of the PSF structures at $25.5\um$ as the maximum flux at a given radial distance. The total flux is scaled to the flux of the $\epsilon$\,Eri stellar photosphere. To scale it to the HD\,107146 stellar photosphere, one has to divide it by 42.3.}
\label{plot_psf}%
\end{figure}

The \emph{accuracy of stellar PSF subtraction necessary} for the different runs is given in Table~\ref{psf_subtraction}. This is the result of a comparison of the simulated disk observations (Fig.\ref{sequ_jwst}) and the results plotted in Fig.~\ref{plot_psf}. We find that a PSF subtraction accuracy of 1\% is sufficient to detect and spatially resolve all simulated debris disks considered in this study (significantly lower accuracy is needed in some cases). Since this uncertainty dominates the total uncertainty in the observations, very accurate photometry is required to scale the reference PSF to the correct level. This represents the minimum requirements to unequivocally detect structures in the model debris disks without further considerations. However, a sophisticated approach might include proper rotation of the optics, so that edge-on seen disks are imaged between the bright hexagonal wings of the PSF -- in regions where the PSF structures are less bright -- increasing the depth of the observations. Furthermore, it might be possible in some cases to scale the reference PSF to the high order structures of the PSF in regions of the science observations where no significant signal from the disk is expected.
\footnotetext{Of the two peaks of the PSF next to $\rho_{\rm disk}$ in radial distance, the brighter one is considered.}

\section{Conclusions}
\label{conc}

We demonstrated that planet-disk interaction may produce detectable structures in the dust distribution of debris disks. This depends on the configuration of the planet-disk system and on the observing wavelength. The detected structures enable one to infer and characterize extrasolar planets in a range of mass and radial distance from the star unattainable by other techniques. In particular, detailed modeling of a combination of high-sensitivity, high spatial resolution observations at mid-infrared wavelengths and \mbox{(sub-)mm} wavelengths is able to put strong constraints on the configuration of the planet-disk system. We demonstrated that HST scattered-light observations are in most cases unable to unambiguously detect such structures, in particular in debris disks seen face-on. In contrast, both ALMA and the JWST will provide the sensitivity and resolution to detect and spatially resolve the spatial dust distribution in debris disks at a level of sensitivity and resolution that allow one to distinguish between different planet-disk configurations. However, we also demonstrated limitations of the instruments. ALMA is unable to reach both high sensitivity to surface brightness and high spatial resolution simultaneously, requiring a sophisticated observing strategy to optimize the outcome of planned observations. For a debris disk with typical shape of the SED, intermediate observing wavelengths around 1\,mm and small to intermediate array extents are ideally suited to reach a high S/N and reasonable spatial resolution with ALMA. The highest sensitivity at reasonable spatial resolution is reached if the resolution of the observations is similar to the scale of the structures expected, as long as these structures are bright enough. Mid-infrared observations of debris disks with the JWST will be able to detect and spatially resolve dust in debris disks even at a distance of several tens of AU, where the emission from debris disks in this wavelength range is expected to be low. For such observations, stellar PSF subtraction with an accuracy of a few percent is necessary to unequivocally detect structures in the spatial distribution of the dust.

\begin{acknowledgements}
We thank J. P. Ruge for valuable discussions and important contributions to the consideration of phase noise. S. Ertel acknowledges financial support from DFG under contract WO\,857/7-1 and from the French National Research Agency (ANR) for financial support through contract ANR-2010 BLAN-0505-01 (EXOZODI), and for general support by K. Ertel.
\end{acknowledgements}

\bibliographystyle{aa}

\bibliography{bibtex}

\begin{appendix}
\section{Orbit integration of mass-less test particles}

\subsection{Equation of motion}

Because no interaction among the dust grains is considered, one can treat them as mass-less test particles independently of each other. Summarizing the different force terms, the equation of motion governing the dynamics of a dust particle at stellocentric position $\textbf{\textit{r}}$ and velocity $\textbf{\textit{v}}$ can be written as \citep{bur79,lio95,mor02}
\begin{eqnarray}
\label{eq_forces}
\textbf{\textit{a}} &=& \textbf{\textit{a}}_{\rm G} + \textbf{\textit{a}}_{\rm R} +
\textbf{\textit{a}}_{\rm PR} + \textbf{\textit{a}}_{\rm SW} + \textbf{\textit{a}}_{\rm Pl}\\[4mm]
&=& - \frac{GM_{\star}(1-\beta)}{r^3}\textbf{\textit{r}} -
\frac{\beta(1+\xi)}{c}\frac{GM_{\star}}{r^2} \left(
\frac{\dot{r}}{r}\textbf{\textit{r}} + \textbf{\textit{v}} \right) \nonumber \\
 & &+ \sum_{j=1}^{P}GM_{{\rm pl},j} \frac{\textbf{\textit{r}}_{{\rm pl},j}-\textbf{\textit{r}}}{|\textbf{\textit{r}}_{{\rm pl},j}-\textbf{\textit{r}}|^3}\\[4mm]
&=& - \frac{GM_{\star}}{r^3} \left \{ (1-\beta) \textbf{\textit{r}} +
\frac{\beta(1+\xi)}{c} \left(
\dot{r}\textbf{\textit{r}} + r \textbf{\textit{v}} \right) \right\} \nonumber \\
 & &+ \sum_{j=1}^{P}GM_{{\rm pl},j} \frac{\textbf{\textit{r}}_{{\rm pl},j}-\textbf{\textit{r}}}{|\textbf{\textit{r}}_{{\rm pl},j}-\textbf{\textit{r}}|^3},
\end{eqnarray}
where $G$ is the gravitational constant, $M_{\star}$ is the stellar mass, $P$ is the number of planets (1 in our case), and $M_{{\rm pl},j}$ and $\textbf{\textit{r}}_{{\rm pl},j}$ are the mass and position vector of planet $j$. The quantity $\beta$ is the ratio of radiation pressure and stellar gravity force \citep{bur79}. The strength of the corpuscular stellar wind (consisting mostly of protons) is prescribed by the parameter $\xi$, defined as the ratio of stellar wind drag to Poynting-Robertson drag. Following \citet{gus94}, we adopt the solar value of 0.35.

\subsection{Orbit integration}

In this section, we only briefly describe the computational procedure; for a complete description we refer the reader to \citet{kok98}. The orbits of the dust particles and that of the planet are integrated using a fourth-order Hermite scheme, a direct integration method using a predict-evaluate-correct (PEC) algorithm to solve the $N$-body problem \citep{mak91a,mak92}. For a particle $i$, the acceleration $\textbf{\textit{a}}_{0,i}$ and its first derivative with respect to time $\dot{\textbf{\textit{a}}}_{0,i}$ are used to predict the future position $\textbf{\textit{r}}_i(t)$ and velocity $\textbf{\textit{v}}_i(t)$ from the instantaneous values $\textbf{\textit{r}}_{0,i}$ and $\textbf{\textit{v}}_{0,i}$ at time $t_0<t$. 

The higher-order derivatives $\textbf{\textit{a}}_{0,i}^{(2)}$ and $\textbf{\textit{a}}_{0,i}^{(3)}$
are obtained directly by Hermite interpolation from $\textbf{\textit{a}}$ and $\dot{\textbf{\textit{a}}}$ at times $t_0$ and $t$ \citep{mak92}. They are used to improve the accuracy of the force polynomial to refine the predicted position and velocity of the particle. For all simulations we iterate the sequence of (1) evaluation of the acceleration and its first derivative and (2) correction of the position and velocity with higher derivates three times to increase the accuracy of the orbit integration.

For the choice of the individual time step, we use the prescription proposed by \citet{aar85}. It uses the acceleration and its first, second, and third time derivatives to compute the time step at which the integration for particle $i$ should be advanced in time. The (continuous) time step values are then 'quantized' to the nearest negative powers of 2 to further economize the computation \citep{mak91b}. Then the computation proceeds in blocks of particles that are integrated simultaneously.

We have tested the accuracy of our numerical integration against the semi-analytical solutions of \citet{wya50} for initially circular and elliptical orbit of a dust particle. The temporal evolution of the orbital elements and the Poynting-Robertson decay time agree very well with \citet{wya50}. The relative energy error of the orbit integration is typically below $10^{-6}$.

\end{appendix}

\end{document}